\documentstyle[11pt,aaspp4,flushrt]{article}
\newcommand{\Ref}{\hangindent=20pt \hangafter=1 \noindent}
\newcommand{\StartRef}{\hyphenpenalty=10000 \raggedright}

\newcommand{\beq}{\begin{equation}}
\newcommand{\eeq}{\end{equation}}
\newcommand{\NarrowMargins}{
  \setlength{\oddsidemargin}{+0.3in}
  \setlength{\evensidemargin}{-0.0in}
  \setlength{\textwidth}{6.2in}
  \setlength{\topmargin}{-0.75in}
  \setlength{\textheight}{9.25in}   }
\catcode`\@=11 % This allows us to modify PLAIN macros.

\def\lsim{\mathrel{\mathpalette\@versim<}}
\def\gsim{\mathrel{\mathpalette\@versim>}}

\def\tt{\times 10^{-3}}
\def\tfo{\times 10^{-4}}
\def\tfi{\times 10^{-5}}
\def\tsi{\times 10^{-6}}
\def\tse{\times 10^{-7}}

\def\mo{\dot m_{\rm out}}
\def\ro{r_{\rm out}}

\def\@versim#1#2{\vcenter{\offinterlineskip
        \ialign{$\m@th#1\hfil##\hfil$\crcr#2\crcr\sim\crcr } }}
\catcode`\@=12 % at signs are no longer letters
\NarrowMargins
\begin{document}
\title{Spectral Models of Advection-Dominated
Accretion Flows with Winds} 
\author{Eliot Quataert\footnote{equataert@cfa.harvard.edu} and Ramesh Narayan\footnote{rnarayan@cfa.harvard.edu}} 
\affil{Harvard-Smithsonian Center for Astrophysics, 60 Garden St., Cambridge, MA 02138}
\medskip
\setcounter{footnote}{0}

\begin{abstract}

We calculate spectral models of advection-dominated accretion flows,
taking into account the possibility that significant mass may be lost
to a wind.  We apply the models to the soft X-ray transient V404 Cyg
in quiescence and the Galactic center source Sgr A*.  We show that
there are qualitative degeneracies between the mass loss rate in the
wind and parameters characterizing the microphysics of the accretion
flow; of particular importance is $\delta$, the fraction of the
turbulent energy which heats the electrons.  For small $\delta$,
current observations of soft X-ray transients and Sgr A* suggest that
at least $\sim 10 \%$ of the mass originating at large radii must
reach the central object.  For large $\delta \sim 0.3$, however,
models with significantly more mass loss are in agreement with the
observations. We also discuss constraints on advection-dominated
accretion flow models imposed by recent radio observations of NGC 4649
and other nearby elliptical galaxies.  We conclude by highlighting
future observations which may clarify the importance of mass loss in
sub-Eddington accretion flows.

\noindent {\em Subject headings:} accretion, accretion disks -- black hole physics 
\end{abstract}

\section{Introduction}

A number of authors have argued that, at sub-Eddington accretion
rates, the gravitational potential energy released by turbulent
stresses in an accretion flow may be stored as thermal energy, rather
than being radiated (Ichimaru 1977; Rees et al. 1982; Narayan \& Yi
1994, 1995; Abramowicz et al. 1995; Chen et al. 1995; see Narayan,
Mahadevan, \& Quataert 1998b, and Kato, Fukue, \& Mineshige 1998 for
reviews).  Narayan \& Yi (1994,1995) noted that such
advection-dominated accretion flows (ADAFs) have the interesting
property that their Bernoulli parameter, a measure of the sum of the
kinetic energy, gravitational potential energy, and enthalpy, is
positive; since, in the absence of viscosity, the Bernoulli parameter
is conserved on streamlines, the gas can, in principle, escape to
``infinity'' with positive energy. Narayan \& Yi speculated that this
might make ADAFs a natural candidate for launching the outflows/jets
seen to originate from a number of accretion systems.

Blandford \& Begelman (1998; hereafter BB98) have recently suggested
that mass loss via winds in ADAFs may be both dynamically crucial and
quite substantial.  They construct self-similar ADAF solutions in
which the mass accretion rate in the flow varies with radius $R$ as
$\dot M \propto R^p$.  If the wind carries away roughly the specific
angular momentum and energy appropriate to the radius from which it is
launched, they show that the remaining (accreting) gas has a negative
Bernoulli parameter only for large values of $p \sim 1$.  They
therefore propose that the majority of the mass originating at large
radii is lost to a wind.  For example, for $p=1$, only a fraction
$\sim(R_{in}/R_{out})\ll1$ of the mass would accrete onto the central
object, where $R_{in}$ and $R_{out}$ are the inner and outer radii of
the ADAF.

In a separate study, Di Matteo et al. (1998; hereafter D98) measured
the flux of radio and submillimeter emission from the nuclei of nearby
elliptical galaxies and found fluxes significantly below the values
predicted by the ADAF model.  Their observations are difficult to
reconcile with Fabian \& Rees's (1995) proposal that these galactic
nuclei contain ADAFs.  D98 discuss a number of explanations for the
``missing'' flux; one of their suggestions is that a significant wind
may carry off much of the accreting mass in the ADAF.

Spectral models of ADAFs without mass loss have been applied to a
number of low luminosity accreting black hole systems.  They give a
satisfying description of the spectral characteristics of several
quiescent black hole binaries (Narayan, McClintock, \& Yi 1996,
Narayan, Barret, \& McClintock 1997; Hameury et al. 1997) and low
luminosity galactic nuclei, e.g., Sgr A* (Narayan, Mahadevan, \& Yi
1995; Manmoto et al. 1997; Narayan et al. 1998) and NGC 4258 (Lasota
et al. 1996a, Gammie, Narayan, \& Blandford 1998).  

Our goal in this paper is to use broad-band spectral observations to
test for the presence of mass loss in low luminosity accreting black
holes, paying special attention to the implications of uncertainties
in the microphysics of the accretion flow.  Specifically, we attempt
to answer the following question: are the no-mass loss ADAF models in
the literature, which fit the observations reasonably well, unique
``ADAF'' fits to the data, or are models with substantial mass loss
also viable?  If the latter, since it is unlikely that purely
theoretical arguments will be definitive, can we distinguish between
no-wind and wind models with future observations?  As a first step
toward addressing these questions, we calculate spectral models of
ADAFs with $\dot M \propto R^p$, and compare them with observations of
the X-ray binary V404 Cyg in quiescence, the Galactic center source
Sgr A*, and the nucleus of the elliptical galaxy NGC 4649.  We assume
throughout that all observed radiation from the systems under
consideration is due to the accretion flow, i.e., the wind/outflow
does not radiate significantly.

In the next section (\S2), we discuss our modeling techniques.  We
then show models for V404 Cyg (\S3) and Sgr A* (\S4) and compare the
models to observations, focusing on the available theoretical
parameter space.  In \S5 we discuss D98's results on the radio
emission in nearby ellipticals.  We then propose several future
observations which may help clarify the physics of ADAFs (\S6).
Finally, in \S7 we summarize and discuss our results.

\section{Modeling Techniques}
Over the last few years, ADAF models have seen a series of
improvements such that the modeling techniques used currently are much
superior to earlier methods.  The first published spectral models of
ADAFs used the self-similar solution of Narayan \& Yi (1994) to model
the dynamics, but this was soon replaced by global models, initially
for a pseudo-Newtonian potential (Narayan, Kato, \& Honma 1997, Chen,
Abramowicz, \& Lasota 1997), and more recently in the full Kerr metric
(Abramowicz et al. 1996, Peitz \& Appl 1997, Gammie \& Popham 1998).
The spectral modeling too has seen improvements, particularly in the
treatment of the electron energy equation and the Comptonization.  The
electron energy equation was originally taken to be local (e.g.,
Narayan, McClintock, \& Yi 1996), with heating due to Coulomb
collisions and turbulent heating balancing cooling.  As emphasized by
Nakamura et al. (1997), however, the electron entropy gradient
(electron advection) generally cannot be neglected, and so this is now
included (see eq. [\ref{ee}]).

Narayan et al. (1998a) discuss how the predicted spectra have changed
as the modeling techniques have improved.  The changes have generally
been fairly modest, at least compared to the large changes we see in
the present paper when we include mass loss from the accretion flow.

In this paper, we use the latest techniques for numerically
calculating spectral models of ADAFs (plus any thin disk at large
radii), as described in detail by Narayan et al. (1998a; see also
Esin, McClintock, \& Narayan 1998 and Narayan, Barret, \& McClintock
1997). Here we mention only the relevant differences.

As in Narayan et al. (1998a) and Esin et al. (1997), we solve the full
electron energy equation, including the electron entropy gradient.
The equation takes the form \beq n_e v {d \over dR } \left({k T_e
\over \gamma_e - 1}\right)= k T_e v {d n_e \over d R}+H_e
+q_{ie}-q_e^-,
\label{ee} \eeq where $T_e$ is the electron temperature, 
$\gamma_e$ is the adiabatic index of the electrons, $n_e$ is the
electron number density, $v$ is the radial velocity, $H_e$ is the
turbulent heating rate of the electrons, $q_{ie}$ is the energy
transferred to the electrons from the ions by Coulomb collisions, and
$q_e^-$ is the radiative cooling rate of the electrons.  The first
term on the right hand side of equation (\ref{ee}) describes the
increase in the electron internal energy due to $PdV$ work, and is the
volumetric version of $q_c$ defined in equation (\ref{comp}) below.

A difference in this paper, relative to earlier work, is that we take
$\gamma_e$ to be that of a monatomic ideal gas ($5/3$ in the
non-relativistic limit, decreasing to $4/3$ in the relativistic
limit).  Esin et al. (1998) argued that $\gamma_e$ should include
contributions from the magnetic energy density in the flow.  As
discussed in Quataert \& Narayan (1998; their Appendix A), this is
incorrect if MHD adequately describes the accretion flow.  This is of
some significance for models of low luminosity systems.  For example,
in the ``standard'' ADAF model of Sgr A*, the electrons are, to good
approximation, adiabatically compressed.  The larger $\gamma_e$ used
in this paper yields higher electron temperatures (by a factor of
$\sim 3$) and significantly more synchrotron emission.  As a result,
to produce a radio flux comparable to that in Narayan et al. (1998a),
we require a noticeably weaker magnetic field.

We describe the turbulent heating of the electrons via a parameter
$\delta$, defined by $H_e \equiv \delta q^+$, where $q^+$ is the usual
``viscous'' dissipation rate of accretion theory (e.g., Kato et
al. 1998).  Thus, $\delta$ is the fraction of the total energy
generated by turbulent stresses in the fluid ($q^+$) that directly
heats the electrons.  As discussed in Quataert \& Narayan (1998),
there is a subtlety in interpreting $q^+$ in ADAFs which is not
present in thin disks; namely, only a fraction $\eta$ ($\sim 1/2$) of
$q^+$ is likely to end up in the particles; the rest is used to build
up the magnetic field and turbulence as the accreting gas flows
in.\footnote{Essentially, the parameter $\eta$ reflects the fact that,
just as one must account for advection by the particles, one must also
account for advection by the turbulence.}  Of the fraction $\eta$, a
fraction $\delta_H$ goes into electrons and $(1-\delta_H)$ goes into
ions.  Thus, in terms of $\eta$ and $\delta_H$, the $\delta$ we use in
this paper is $\delta \equiv \delta_H\eta$.

Accounting for a variable mass accretion rate in the flow, the
continuity equation becomes \beq \dot M = - 4 \pi R^2 H_\theta \rho v
= \dot M_{\rm out} \left({R \over R_{\rm out}}\right)^p, \label{cont}
\eeq where $H_\theta$, $\rho$, and $v$, are, respectively, the angular
scale height, mass density, and radial velocity in the flow.  The
quantity $\dot M_{\rm out}$ is the accretion rate at the radius
$R_{\rm out}$, where winds become important.  We take the radial
velocity, angular velocity, and sound speed of the flow from the
global, relativistic, models of Gammie \& Popham (1998), and then use
equation (\ref{cont}) to calculate the density, $\rho$.  This is,
strictly speaking, inconsistent, as Gammie \& Popham's models were
derived under the assumption of constant $\dot M$.  The error made in
this approximation should, however, be of order unity.  From a
spectral modeling point of view, the primary importance of the wind is
that it modifies the density in the flow; this is correctly captured
by equation (\ref{cont}).

Generically, ADAFs with winds will rotate more quickly than those
without winds.  This is seen in the self-similar solution of BB98,
where the rotational support enables the enthalpy of the gas to
decrease, thus permitting the Bernoulli parameter to become negative.
The shear and the viscous dissipation per unit mass in the flow are
therefore expected to be larger in the presence of a
wind.\footnote{This is actually true only for certain ``types'' of
winds (in particular, depending on BB98's parameters $\epsilon$ and
$\lambda$).  As discussed in, e.g., Blandford \& Payne (1982), it is
possible for a wind to take away all of the angular momentum and
energy flux from the disk, leaving it cold and dissipationless.}  We
have crudely accounted for this effect as follows.

In non-wind models, the flow structure is, among other things, a
function of $\gamma_g$, the adiabatic index of the fluid.  In
calculating models of systems with winds and high $\delta$ (Figures
2b, 4b, 7, \& 8), we have chosen $\gamma_g$ such that it yields a
rotation rate in the interior of the flow which is comparable to that
expected from the self-similar wind solution of BB98.  In particular,
a self-similar, non-relativistic, ADAF has a Bernoulli parameter equal
to zero only if $\Omega/\Omega_K \approx [2p/(p + 5/2)]^{1/2}$.  For
our typical value of $p = 0.4$, this yields $\Omega/\Omega_K \approx
0.53$.  In this case we take $\gamma_g \approx 1.5$ in our global
calculations, since it reproduces this rotation rate well.  Note that
it is important to get the right $\Omega$ only for high $\delta$.  For
low $\delta$, since turbulent heating of electrons is unimportant, the
exact $\gamma_g$ we use is not important.  We have confirmed this by
calculating models with various choices of $\gamma_g$ at low $\delta$
and making sure that the spectral models are only weakly modified.

We are reasonably confident that, even though we have used an ad hoc
prescription in choosing the global solutions, our parameter estimates
are fairly accurate.  Ultimately, of course, global, relativistic,
models of ADAFs with winds will be needed to correctly assess some of
the issues addressed in this paper.

\subsection{Choice of parameters}

We measure black hole masses in solar units and (radially varying)
accretion rates in Eddington units: $M = m M_{\odot}$ and $\dot M =
\dot m \dot M_{edd}$.  We take $\dot M_{edd} = 10L_{edd}/c^2 = 2.2
\times 10^{-8} m M_{\odot} {\rm yr}^{-1}$, i.e., with a canonical 10
\% efficiency.  We measure radii in the flow in Schwarzschild units:
$R = r R_s$, where $R_s = 2GM/c^2$ is the Schwarzschild radius of the
black hole.

The parameters of our models are $m$, $\mo = \dot M_{\rm out}/\dot
M_{\rm edd}$, $\beta$, $\alpha$, $\delta$, $\ro$, and $p$.  Our
primary focus is to consider the effects of winds via the parameter
$p$ (defined in eq. [\ref{cont}]).  As we show, however, variations in
$p$ are qualitatively degenerate with variations in other parameters
of the problem.

The mass of the central black hole, $m$, is estimated from
observations.  As in all previous work, we fix $\mo$ by adjusting it
so that the X-ray flux in the model fits the available data.  For all
of the models presented here, $\ro=10^4$.  Note that $\ro$ and $p$
are, roughly speaking, degenerate; what is of primary importance is
$\ro^{-p}$, the fraction of the incoming mass accreted onto the
central object.  Typical values of $p$ considered are $p = 0$ (no
winds) and $p = 0.4$ (moderately strong wind).  

The quantities $\beta \equiv P_{\rm gas}/P_{\rm mag}$, $\alpha$, and
$\delta$ are microphysical parameters representing the magnetic field
strength in the flow, the efficiency of angular momentum transport,
and the fraction of the turbulent energy which heats the electrons,
respectively.\footnote{Our definition of $\beta$ is that utilized in
the plasma physics literature.  A number of workers in the accretion
literature define a ``$\beta$'' via $\beta_{\rm adv} \equiv
P_{gas}/P_{tot}$, with $P_{tot} = P_{gas} + P_{mag}$.  This is related
to our $\beta$ by $\beta_{\rm adv} = 3\beta/(3\beta + 1)$ or
$\beta_{\rm adv} = \beta/(\beta + 1)$, depending on whether one
defines the magnetic pressure to be $B^2/24 \pi$ or $B^2/8 \pi$ (as we
do here).}  Recent ADAF models in the literature have favored the
values $\beta = 1$, $\alpha = 0.25$, and $\delta = 10^{-3}$
(cf. Narayan et al. 1998a), and have considered only factor of few
variations in $\alpha$ and $\beta$ and factor of $\sim 10$ variations
in $\delta$.  There is, however, considerable uncertainty in the
microphysics of ADAFs; each of the above parameters must be regarded
as uncertain to at least an order of magnitude, likely more.

As we will show in this paper, mass loss from the accretion flow has a
dramatic effect on theoretically predicted spectra.  If we were to
restrict ourselves to the values of the microphysics parameters given
above, significant mass loss would be all but ruled out by the
observations.  Such a restriction would, however, be an inaccurate
reflection of the theoretical uncertainty in the microphysics of the
flow.  The philosophy adopted in this paper is therefore somewhat
different from previous studies.  We allow $\alpha$, $\beta$ and
$\delta$ to vary over a much larger range, but one which we believe
correctly encompasses the theoretical uncertainties.

For purely theoretical reasons (see below) we take our ``canonical''
values to be different from those of previous studies, namely, $\beta
= 10$, $\alpha = 0.1$, and $\delta = 10^{-2}$.  By canonical we mean
(only) that, when one of the parameters is varied (e.g., $p$), the
others (e.g., $\delta$) are typically fixed at their canonical values.
A major point of this paper will be that, depending on the importance
of winds, these values may or may not be consistent with observations.

Theoretical work on particle heating in ADAFs (Gruzinov 1998, Quataert
1998, Quataert \& Gruzinov 1998) and ``fluid'' models for the
evolution of the turbulent energy in an ADAF (Quataert \& Narayan
1998) suggest that subthermal magnetic fields may be likely; we
consider $\beta = 10$ to be a plausible value.  We take, however, a
range of $\beta$, from $\beta=1$ (strict equipartition of gas and
magnetic pressure) to $\beta=100$ (weak fields).

If the turbulent stresses arise solely from magnetic fields, we expect
the viscosity parameter to scale roughly as $\alpha\sim1/\beta$
(Hawley, Gammie \& Balbus 1996).  We do not always enforce this
relation in our models, but sometimes vary $\alpha$ and $\beta$
independently.  We consider values of $\alpha$ ranging from 0.03 to
0.3.  We should note, however, that large values of $\alpha \approx
0.25$ are needed in applications of the ADAF model to X-ray binaries
such as Nova Muscae 1991, Cyg X--1 and GRO J0422+20 in the low/hard
state (Narayan 1996, Esin, McClintock \& Narayan 1997, Esin et
al. 1998).  If $\alpha$ is much smaller than 0.25, the maximum
accretion rate, $\dot m_{crit}$, up to which the ADAF solution is
possible decreases significantly, and the maximum luminosity of the
models becomes much smaller than the observed luminosities.  We have
confirmed that this limit on $\alpha$ is not modified if winds are
included in the models.

The value of the parameter $\delta$ is uncertain.  Traditional ADAF
models have taken $\delta$ to be small ($\sim 10^{-3})$ and never
considered $\delta \gsim 0.03$.  A number of studies have been carried
out to investigate the heating of protons and electrons in hot
plasmas.  Quataert (1998) and Gruzinov (1998; see also Blackman 1998,
Quataert \& Gruzinov 1998) considered particle heating by MHD
turbulence and concluded that $\delta$ might be small so long as
$\beta$ is greater than about $\sim 10$.  Bisnovatyi-Kogan \& Lovelace
(1997; see also Quataert \& Gruzinov 1998), however, argue that
magnetic reconnection, and its presumed electron heating, may lead to
large values of $\delta \sim 1$.\footnote{Despite our disagreement
with some of Bisnovatyi-Kogan \& Lovelace's arguments (see Blackman
1998), their basic point, that reconnection may be crucial for ADAF
models, is nonetheless important.} In this paper, we avoid theoretical
prejudice and consider values of $\delta$ ranging from 0 to 0.75.
Since the maximum value of $\delta$ is $\eta$, the fraction of the
turbulent energy that goes into the particles, the value $\delta =
0.75$ likely corresponds to a situation where electrons are heated
much more strongly than ions.

\subsection{Description of Spectra}

In the following sections we compare theoretical spectra of ADAFs,
both with and without winds, to observations of low-luminosity
systems.  In preparation for this we introduce here the main features
of the calculated spectra.

Three radiation processes are of importance in ADAF spectra:
synchrotron emission, Compton scattering, and bremsstrahlung.  Each of
these produces distinct and easily recognized features in the
spectrum.  The relative importance of each mechanism is a function of
the temperature and density of the plasma, and thus of the model
parameters, $\alpha$, $\beta$, $\delta$, $p$, $m$, and $\dot m$.

Thermal synchrotron emission in ADAFs is invariably self-absorbed and
produces a sharply cutoff peak, with a peak frequency that depends on
the mass of the black hole: $\nu_{s} \sim10^{15}m^{-1/2}$ Hz.  The
synchrotron peak is in the optical band for stellar-mass black holes,
and in the radio for supermassive black holes.  Synchrotron emission
from different radii in the flow occurs at different frequencies.  The
peak emission, however, is always from close to the black hole and
reflects the properties of the accreting gas near $r \sim 1$. 

In spectra of quiescent systems of the sort we discuss in this paper
($\dot m_{in}\lsim10^{-3}$), and especially in the absence of winds,
the synchrotron peak is the most luminous feature in the spectrum.
The maximum value of $\nu L_\nu$ is given by (Mahadevan 1997) \beq
\nu_s L_{\nu,s} \propto B^3 T^7_e \propto \beta^{-3/2} \dot
m_{in}^{3/2} T^7_e, \label{lsynch} \eeq where all quantities should be
evaluated at $r \sim 1$ and $\dot m_{in}$ is the accretion rate near
$r \sim 1$.  Note the very steep dependence on the electron
temperature.

In writing equation (\ref{lsynch}), we have taken $\rho \propto \dot
m$, but independent of $\alpha$, as is appropriate near the central
object.  This can be understood by noting that, near the central
object, the self-similar scaling $\rho \propto \dot m/\alpha$ fails.
The dynamics in the synchrotron and Compton emitting regimes is
dominated by the presence of a sonic point at $r \sim 3-5$ (Narayan,
Kato, \& Honma 1997, Chen et al. 1997, Gammie \& Popham 1998).  Near
this radius the flow velocity is $\sim$ the sound speed, independent
of $\alpha$.  By the continuity equation, then, the density in the
interior scales as $\rho \propto \dot m$, with only a weak dependence
on $\alpha$.  The density on the outside, however, does scale as $\rho
\propto \dot m/\alpha$ because, away from the sonic point,
self-similarity is reasonably valid.

Compton scattering of synchrotron photons by the hot electrons in the
accreting gas produces one, or sometimes two peaks in the spectrum at
frequencies higher than the synchrotron peak.  The peaks correspond
to successive scatterings by the electrons.  As with the synchrotron
peak, the Compton features are again sensitive to the properties
of the gas near the black hole.

The frequency of the first Compton peak $\nu_{c}$ is related to
$\nu_{s}$ by the Compton boost factor $A$, which is a function only of
the electron temperature \beq
{\nu_{c}\over\nu_{s}}=A=1+4\theta_e+16\theta_e^2, \qquad
\theta_e={kT_e\over m_ec^2}={T_e\over 5.9\times10^9~{\rm
K}}. \label{A} \eeq The power in the Compton peak relative to that in
the synchrotron peak depends on both $A$ and the optical depth of the
flow to electron scattering ($\tau$) \beq \nu_c L_{\nu,c} \approx
\nu_s L_{\nu,s} \left({\nu_c \over \nu_s}\right)^{\alpha_c}, \qquad
\alpha_c = 1 + {\ln \tau \over \ln A}. \label{compt} \eeq The relative
power in the synchrotron and Compton peaks therefore provides some
information on the density in the inner regions of the flow, and thus
on $\dot m_{in}$.  Note that for the low luminosity systems considered
here, $\tau \ll 1$ and $\alpha_c < 0$, so that the synchrotron
luminosity dominates the Compton luminosity.

Finally, bremsstrahlung emission produces a peak that typically
extends from a few to a few hundred keV.  In contrast to the other two
processes discussed above, this emission arises from all radii in the
flow.  To see this, consider a self-similar ADAF with a wind, for
which $\rho \propto r^{-3/2 + p}$ and $T_e \propto r^{-\epsilon}$
($\epsilon \sim 1$ at large radii).  Let the minimum flow temperature be
$T_{min}$ (which occurs at $r_{out}$) and the maximum temperature be
$T_{max}$ (near $r\sim1$).  At photon energies $\ll kT_{min}$, the
bremsstrahlung emission is given roughly by $\nu L_\nu \propto \nu$
(the spectral index is a little different from unity because of the
Gaunt factor, which we ignore for simplicity), while for $kT_{min} \ll
h \nu \ll kT_{max}$, it is $\nu L_\nu \propto \nu^{1/2 - 2p/\epsilon}$.
In each case, the emission comes from the largest radius which
satisfies $h \nu \sim k T(r)$.  In our models, $T_{min}$ is $\sim
(10^{12}/\ro)$ K; therefore, for $\ro=10^4$, $kT_{min}$ is $\sim
10$keV.  For X-ray observations in the range $0.1-10$keV, $h \nu \lsim
kT_{min}$.  By above, $\nu L_{\nu}$ should be roughly proportional to
$\nu$, but should flatten beyond about 10 keV; $\nu L_\nu$ will vary
as $\nu^{1/2}$ beyond 10 keV if there is no wind ($p = 0$) and it will
be flatter or even turn over (in $\nu L_\nu$) if there is a strong
wind (large $p$).

In all cases, the hardest emission, at $\gsim 100$ keV, occurs from
the inner $r \lsim 100$, while the softer emission comes from $\sim
\ro$ (this is particularly true for $p > 0$).  Observations in the
1--10 keV X-ray band are therefore most sensitive to the outer regions
of the ADAF.  In these regions, the electron temperature is fairly
well-determined since the gas is essentially virial and
one-temperature.  Therefore, observations of the bremsstrahlung
emission at a few keV give direct information on the density of the
outer flow, and thereby the accretion rate on the outside $\dot
m_{out}$.

In the sources that we consider below, the synchrotron peak is
isolated and well observed (in X-ray binaries, the companion must be
subtracted out); it occurs in softer bands, either in the optical or
radio.  The Compton and bremsstrahlung peaks, however, can sometimes
be superposed in the X-ray band.  In particular, an important
consequence of the $m$ dependence of the frequency of the synchrotron
peak ($\nu_s$) is that, without winds, the X-ray spectrum of ADAF
models of low luminosity galactic black hole candidates is usually
dominated by the first Compton peak.  In low luminosity AGN, however,
the precise behavior in the X-ray band is sensitive to the details
(microphysics, accretion rate) of the model, being a competition
between the second Compton peak and bremsstrahlung.  This is because
the peak synchrotron emission is at substantially lower frequencies
and a synchrotron photon must be scattered more than once (or off of
hotter electrons) in order to be scattered into the X-ray band; this
tends to suppress the importance of Comptonization.  Note that
bremsstrahlung and Comptonization can be readily distinguished by
their different spectral slopes in the X-ray band.

If there is a thin disk outside the ADAF, as in our models of X-ray
binaries (\S3), the emission of the disk is seen as a blackbody-like
feature in the spectrum.  This emission is in the red or near infrared
for quiescent X-ray binaries in which the disk is restricted to
$r>\ro\sim10^4$.

\subsection{The Effects of Winds on Spectral Models}

Bremsstrahlung emission at $\sim 1-10$ keV is rather insensitive to
the presence of a wind (i.e., to $p$) since it originates in the outer
regions of the flow and essentially measures $\mo$.  At higher
energies, $\gsim 10$ keV, however, the bremsstrahlung emission
decreases with increasing $p$ ($\nu L_\nu \propto \nu^{1/2 -
2p/\epsilon}$) and thus provides a powerful probe of the flow density
and the value of the parameter $p$ (see \S 6.2).

By contrast, the predicted synchrotron emission decreases strongly
with increasing $p$.  There are two reasons for this.  First,
increasing $p$ decreases the density of the plasma near $r \sim 1$,
where the high frequency synchrotron emission originates.  This
implies a lower gas pressure and hence a weaker magnetic field (for
fixed $\beta$).  Perhaps more importantly, the electron temperature
decreases as $p$ increases.  For the low luminosity systems considered
in this paper, and for small $\delta$, the electrons are nearly
adiabatic, i.e., $T_e \propto \rho^{\gamma_e - 1} \propto r^{(-1.5 +
p)(\gamma_e -1)}$.  When $p$ is large, the density profile is flatter,
adiabatic compression is less efficient, and hence $T_e$ is
smaller. By equation (\ref{lsynch}), the synchrotron emission is
particularly sensitive to the electron temperature.  Therefore, the
synchrotron emission falls very rapidly with increasing $p$.  This
effect can, as we show explicitly below, be countered by increasing
$\delta$, since a larger $\delta$ means stronger turbulent heating of
the electrons and thus larger $T_e$.

The Compton power decreases with increasing $p$ even more strongly
than the synchrotron does.  As equation (\ref{compt}) shows, $\nu_c
L_{\nu,c}$ depends on both $\nu_s L_{\nu,s}$ and $\alpha_c$, both of
which decrease because of the wind ($\alpha_c$ decreases because
$\tau$ and $\theta_e$ both decrease).  Increasing $\delta$ to restore
the synchrotron power also increases the Compton power, as discussed
in the following sections.

\section{Soft X-ray Transients in Quiescence}

Soft X-ray transients (SXTs) are mass transfer binaries which
occasionally enter a high luminosity, ``outburst,'' phase, but most of
the time remain in a very low luminosity, ``quiescent,'' phase.  The
spectra of quiescent SXTs are not consistent with a thin accretion
disk model, which is unable to account for the fluxes and spectral
slopes in the optical and X-ray bands consistently (e.g., McClintock
et al. 1995).  Narayan, McClintock, \& Yi (1996; see also Narayan et
al. 1997a, Hameury et al. 1997) showed that this problem can be
resolved if quiescent SXTs accrete primarily via ADAFs, with the thin
disk confined to large radii, $r>r_{\rm out} \sim 10^4$.  (Note that
$r_{out}$ is taken here to be the same as the transition radius,
$r_{tr}$, defined in previous papers.  However, it need not be if
winds only become important well inside the outer boundary of the
ADAF.)

In this section, we give a detailed description of models of the SXT
V404 Cyg in quiescence.  The X-ray data on V404 Cyg (Narayan et
al. 1997a) are much superior to the data on other SXTs, which makes
this system better suited for the parameter study we present.

Table 1 lists the various parameter combinations we have tried for
modeling V404 Cyg, and some of the characteristics of these models,
including the microphysical parameters, the maximum electron
temperature, and the radiative efficiency.  Following Shahbaz et
al. (1994), we have taken the mass of the black hole to be $m = 12$.

Outbursts in SXTs are believed to be triggered by a thermal-viscous
instability in the thin disk (enhanced mass transfer from the
companion may also be important).  Initial ADAF models of black hole
SXTs in quiescence (Narayan, McClintock, \& Yi 1996) assumed that the
observed optical emission from these systems was blackbody emission
from a steady state outer thin disk.  Wheeler (1996) and Lasota,
Narayan, \& Yi (1996) pointed out that this was inconsistent because
quiescent thin disks are not likely to be in steady state.
Furthermore, in non-steady quiescent disks, the mass accretion rate
decreases rapidly with radius (so as to maintain a roughly constant
effective temperature $\sim$ a few thousand K; e.g., Cannizzo 1993).
This implies a limit on $\ro$; if $\ro$ is too small, then the disk
cannot supply sufficient mass to the inner ADAF to fit the X-ray
observations.  Quantitatively, the limit is $\ro \gsim 10^4$ for V404
Cyg (it is slightly smaller for A0620-00).  We fix $\ro = 10^4$ in all
the models presented here, and we take the thin disk to extend from $r
= 10^4$ to $10^5$. 

\subsection{Spectral Models of V404 Cyg}

Figure 1a shows spectral models of V404 Cyg for different $p$ for our
standard microphysics parameters: $\alpha = 0.1$, $\beta = 10$, and
$\delta = 0.01$.  We see two important effects of changing $p$.

First, in the X-ray band, models with weak winds (small $p$) have
Compton-dominated X-ray spectra, while models with strong winds (large
$p$) are bremsstrahlung dominated.  The reason for this has already
been explained in \S2.  The Compton emission comes from near the black
hole, while the bremsstrahlung comes from the outer regions of the
ADAF.  As the wind becomes stronger, the inner mass accretion rate
$\dot m_{in}=\dot m_{out}\ro^{-p}$ becomes significantly smaller than
$\dot m_{out}$, reducing the importance of Comptonization relative to
bremsstrahlung.

Associated with this switch is another interesting feature.  For weak
winds (small values of $p$), we see that $\dot m_{in}$ remains roughly
constant when we change $p$ (e.g. $\dot m_{in}= 10^{-3}, 9 \times
10^{-4}$, for $p=0$, 0.2), while for large values of $p$, it is $\dot
m_{out}$ that remains roughly constant (e.g. $\dot m_{out}=0.016$,
0.02, for $p=0.4$, 0.6).  This is again easy to understand once we
realize that the mass accretion rate is adjusted so as to reproduce
the X-ray flux.  When the model is Compton-dominated, the X-ray flux
depends on $\dot m_{in}$, and so this quantity remains roughly the
same as $p$ varies.  However, when bremsstrahlung dominates, the X-ray
flux depends on $\dot m_{out}$ and so it is $\dot m_{out}$ that
remains constant.

The second effect that is seen in Figure 1a (and even more clearly in
Fig. 4a for Sgr A$^*$) is that the synchrotron emission becomes weaker
as the wind becomes stronger.  Once $p$ is large enough ($\gsim0.2$)
for the model to become bremsstrahlung-dominated, $\dot m_{out}$ is
more or less frozen at a fixed value.  For yet larger $p$, $\dot
m_{in}$ decreases rapidly with increasing $p$.  Since the synchrotron
emission depends primarily on $\dot m_{in}$, the synchrotron peak
drops significantly in magnitude.  The decrease in the synchrotron
power at large $p$ is actually more dramatic than is apparent in
Figure 1a (see Fig. 4a).  Most of the optical/infrared flux in the
$p=0.4$, 0.6 models in Figure 1a is blackbody emission from the outer
disk, which depends only on $\dot m_{out}$, and does not change with
$p$ at large $p$.  This emission is cool (it is limited by the disk's
effective temperature, which is about 5000 K) and the peak occurs at
lower frequencies.

The above analysis hinges on the change in the flow density with $p$.
How is it modified if the microphysical parameters are varied from the
canonical values taken above?  Figure 1b shows models with a
moderate wind, $p = 0.4$, for various values of the parameter
$\beta$, which determines the strength of the magnetic field ($\alpha$
and $\delta$ are fixed at their canonical values of 0.1 and 0.01,
respectively).  Changing $\beta$ has little effect in the X-ray band
since bremsstrahlung emission does not depend on the magnetic field
strength.  Increasing $\beta$ to $\sim 1$ naturally increases the
synchrotron flux (eq. [\ref{lsynch}]).  Even for $\beta = 1$, however,
the synchrotron luminosity is too low by a factor of $\sim 2-3$.  Note
that, for $\beta = 1$, the optical emission in the models of Figure 1b
is primarily synchrotron, while for $\beta \gsim 10$ it is primarily
disk emission.

Figure 2a shows models of V404 Cyg for $p = 0.4$ for several $\alpha$
($\beta$ and $\delta$ are fixed at their canonical values of 10 and
$10^{-2}$).  These models show little variation in X-ray behavior with
$\alpha$, but there is a decrease in optical emission as $\alpha$
decreases.  This can be understood as follows.  In the self-similar
regime (reasonably valid at large radii), the flow density in an ADAF
is $\rho\propto {\dot m}/\alpha$.  Furthermore, the X-ray flux, which
we fix to the observed value, arises from the outer regions of the
ADAF via bremsstrahlung.  Since the bremsstrahlung luminosity is
$\propto \rho^2$, $\mo/\alpha$ remains roughly constant as $\alpha$
varies ($\mo = 5 \times 10^{-3}, 1.6 \times 10^{-2}$, $2.6 \times
10^{-2}$ for $\alpha = 0.03, 0.1$, $0.3$).  All three models therefore
have nearly the same density and temperature on the outside, which
accounts for the lack of significant change in the X-ray band.  For
these models, however, the optical is dominated by disk emission,
which is proportional to $\dot m_{out}$, rather than $\dot
m_{out}/\alpha$.  For smaller $\alpha$, $\mo$ is smaller and thus the
disk emission decreases, as seen in Figure 2a.

Finally, Figure 2b shows models of V404 Cyg for $p = 0.4$ for several
$\delta$ ($\beta = 10$, $\alpha = 0.1$).  For small
$\delta\lsim10^{-2}$, the electrons are heated primarily by adiabatic
compression (the first term on the right in eq. [\ref{ee}]) and so the
results are nearly independent of the value of $\delta$.  However,
once $\delta\gsim10^{-2}$, turbulent heating ($H_e$) becomes the
dominant heating mechanism.  In this regime, increasing $\delta$
causes the electrons to become hotter (see Table 1), thereby
increasing the synchrotron emission and Comptonization.  For
sufficiently large $\delta \gsim 0.1$, Comptonization dominates
bremsstrahlung in the X-ray band, and the spectra begin to resemble
the no-wind model shown by the solid line in Figure 1a.

The above results are for ADAF models with winds, since that is the
primary focus of this paper.  For completeness, we have considered the
sensitivity of no-wind (or weak wind) models to variations in $\alpha,
~\beta$ and $\delta$.  Figure 3a shows models of V404 Cyg with $p = 0$
taking, for brevity, $\alpha \sim \beta^{-1}$, as suggested by
numerical simulations of thin accretion disks (Hawley, Gammie, \&
Balbus 1996).  We see that larger values of $\alpha$ (and lower
$\beta$) lead to more synchrotron emission.  Figure 3b shows models
for various $\delta$.  Increasing $\delta$ leads to a noticeable
increase in the electron temperature.  This is seen explicitly in
Table 1 and also in the larger ``displacement'' of the Compton peak
relative to the synchrotron peak (see eq. [\ref{A}] for the Compton A
parameter).  Since the synchrotron and Compton emission increase
strongly with temperature, the model with the largest $\delta$ has a
significantly lower $\dot m$ (Table 1).

\subsection{Comparison with Observations}

Figures 1-3 show the available observational constraints on the
spectrum of V404 Cyg (taken from Narayan et al. 1997a).  The optical
data give the luminosity of the source and constrain the effective
temperature of the radiation to be $\gsim10^4$ K.  There is an upper
limit on the EUV flux, which is not very interesting since it is
easily satisfied by all the models considered here.  Thanks to an
excellent ASCA observation, the luminosity in the X-ray band is known
accurately, and the spectral index is also well constrained; in terms
of $\nu L_{\nu} \propto \nu^{2-\Gamma}$, the 2 $\sigma$ error bars on
the photon index $\Gamma$ are $2.1^{+0.5}_{-0.3}$.

The observations give a few important constraints.  First, the $>10^4$
K temperature of the optical argues against the outer thin disk as the
source of this radiation (Lasota, Narayan, \& Yi 1996b; see below).
Thus, the optical has to come from synchrotron and this emission must
be stronger than the disk emission.  Second, the observed photon index
in X-rays in V404 Cyg is incompatible with the $\Gamma = 1$ expected
for thermal bremsstrahlung (\S2).  This means that the X-ray emission
has to be Compton-dominated.  There is preliminary evidence that the
same is also true for A0620-00 (Narayan et al. 1996, 1997a), but the
ROSAT data on that source (McClintock et al. 1995) are not
sufficiently good to trust this conclusion; on the other hand, for GRO
J1655-40, preliminary ASCA data in quiescence indicate a much harder
X-ray spectrum than in V404 Cyg and A0620-00 (Hameury et al. 1997).
Finally, the data show that the optical emission is about an order of
magnitude larger (in $\nu L_\nu$) than the X-ray flux, another
constraint that has to be satisfied by models.

The baseline no-wind ($p=0$) model of V404 Cyg, with canonical values
for the microphysics parameters, is shown by the solid line in Figure
1a.  This model fits the observations well, as emphasized by Narayan
et al. (1997a).  It has roughly the right luminosity and effective
temperature in the optical and is consistent with the X-ray data.  The
model shown here differs somewhat in the X-ray band from that shown in
Narayan et al. (1997a).  The difference is due to the different energy
equation used here, which leads to hotter electrons and more
pronounced Compton bumps.  The value of $\dot m$ is also lower by a
factor of a few.

The observed X-ray spectral index in V404 Cyg places interesting
constraints on models.  For weak winds ($p \sim 0$) the models are in
agreement with the observed slope for a wide range of microphysical
parameters (Figure 3).  For strong winds, however, most of the models
are too bremsstrahlung-dominated to fit the X-ray slope.  For small
$\delta \sim 10^{-2}$, the observed slope rules out $p \gsim 0.3$, for
any $\alpha$ and $\beta$ (Figures 1-2).  For the value of $\ro = 10^4$
we have taken, this means that at least $\sim 10 \%$ of the mass
supplied from the companion must reach the central object.

As discussed by Lasota et al. (1996b), a thin disk cannot account for
the observed optical emission in quiescent SXTs.  This is because thin
disk annuli with effective temperatures comparable to the observed
values, $\sim 10^4$ K, are thermally and viscously unstable.  In fact,
within the context of the disk instability model, quiescent disks in
black hole SXTs have effective temperatures $\sim 3000-5000$ K (Lasota
et al. 1996b), too low to account for the observations.  This is an
independent argument against high $p$, low $\delta$ ADAF models, since
the optical emission in these models is always dominated by the disk
(the synchrotron being strongly suppressed by the large $p$).

Perhaps the most interesting result to come out of these comparisons
is that wind models agree with the data for larger values of the
electron heating parameter $\delta$.  The $p=0.4$, $\delta=0.3$ model
in Figure 2b is as good as the no wind low-$\delta$ model shown in
Figure 1a.  The increase in $T_e$ associated with increasing $\delta$
brings the synchrotron emission into rough agreement with the observed
optical flux, despite the low value of $\dot m_{in}$;\footnote{The
synchrotron peak is a little too cool to fit the data; given the model
uncertainties, however, the difference is not large enough to argue
against these models.}  at the same time, it shifts the balance in the
X-ray band from bremsstrahlung to Comptonization, as required by
observations.

\section{The Galactic Center}

Observations of the Galactic Center indicate that the mass of the
black hole in Sgr A* is $m \sim (2.5 \pm 0.4) \times 10^6$ (Haller et
al. 1996; Eckart \& Genzel 1997; Genzel et al. 1997).  The accretion
rate is estimated to lie in the range $10^{-4} \lsim \mo \lsim {\rm
few}\,\times 10^{-3}$ (Genzel et al. 1994; Melia 1992), with the upper
end of the range considered more likely (Coker \& Melia 1997).  For a
radiative efficiency of 10\%, and assuming that $\dot m$ is constant
in the accretion flow, the implied luminosity is between $\sim 10^{40}
$erg s$^{-1}$ and $\sim 10^{42}$erg s$^{-1}$.  This is well above the
bolometric luminosity of $\lsim 10^{37}$erg s$^{-1}$ inferred from
observations in the radio to $\gamma$--rays (see Narayan et al. 1998a
for a review of the observations).

An optically thin, two temperature, ADAF model is a possible
explanation for the low luminosity of Sgr A* (Rees 1982; Narayan, Yi,
\& Mahadevan 1995, Manmoto et al. 1997, Narayan et al. 1998a,
Mahadevan 1998).  An alternative explanation is that most of the gas
supplied at large radii is lost to a wind and very little reaches the
central black hole (BB98).  We consider both
possibilities in this section.

There is little observational evidence in Sgr A$^*$ for (or against) a
particular value of $\ro$.  In addition, there is little evidence that
the accretion outside $\ro$ occurs via a thin disk.  In our models, we
set $\ro = 10^4$ and assume that, whatever form the plasma takes at
larger radii, it is non-radiating.

\subsection{Spectral Models of Sgr A$^*$}

The parameters of each of our models of Sgr A* are given in Table 2.

Figure 4a shows spectral models of Sgr A* for various $p$, taking
$\alpha = 0.1$, $\beta = 10$, and $\delta = 0.01$.  As usual, the
value of $\dot m_{out}$ in each model has been adjusted to fit the
X-ray flux (even though the ROSAT measurement used in the fits is
really only an upper limit; cf. Narayan et al. 1998a).  The results in
Figure 4a are similar to those shown in Figure 1a for V404 Cyg, but the
effects are somewhat more pronounced.

At $\sim 1$ keV, the baseline no-wind ($p=0$) model in Figure 4a
corresponds to an interesting situation: there are roughly equal
contributions from the second Compton bump and bremsstrahlung.  Recall
that increasing $p$ always shifts the balance in favor of
bremsstrahlung.  Therefore, once $p$ is increased above zero, the
Compton flux decreases, and the spectrum becomes
bremsstrahlung-dominated in the X-ray band.  This switch is evident
already at $p=0.2$ and it becomes more pronounced for larger $p$.  The
three bremsstrahlung-dominated models with $p=0.2$, 0.4 and 0.6 all
have nearly the same value of $\dot m_{out} \approx 2 \times 10^{-4}$,
while there is a modest change in $\dot m_{out}$ between $p=0$ and 0.2
(see Table 2).

Another effect seen very clearly in Figure 4a is the decrease in the
synchrotron emission in the radio with increasing $p$.  This is due to
a decrease in both the magnetic field strength and $T_e$ (\S2).  Note,
in particular, that $T_e$ decreases by a factor of $\approx 5$ from $p
= 0$ to $p = 0.6$ (Table 2).  

The dependence of wind models of Sgr A* on the microphysical
parameters is very similar to that of V404 Cyg.  The one exception is
that all of the $p \gsim 0.2$ models in Figure 4-6 are
bremsstrahlung-dominated in X-rays; we practically never see
Comptonized power in the X-ray band.  This is simply because the
source of soft photons -- the synchrotron peak -- is at substantially
lower frequencies in Sgr A* compared to the SXTs (recall that $\nu_s
\propto m^{-1/2}$; \S2.2).

Figures 4b, 5a and 5b show models of Sgr A* for $p = 0.4$ and
different values of $\beta$, $\alpha$ and $\delta$.  We reach two
conclusions from these calculations.  First, no combination of
$\alpha$ and $\beta$ alone is sufficient to bring the synchrotron
emission of wind models back to the level seen in the baseline no-wind
model (Figures 4b \& 5a).  Just as in V404 Cyg, however, increasing
$\delta$ has a very strong effect on the synchrotron emission.
Indeed, a $p=0.4$, $\delta=0.3$ model has roughly the same synchrotron
power as the $p=0$, $\delta=0.01$ no-wind model.  The reason is clear
--- increasing $\delta$ causes a substantial increase in $T_e$ (Table
2), which compensates for the reduced density and field strength due
to the wind.  By contrast, neither $\alpha$ nor $\beta$ has a
comparable effect.

As Figure 5a shows, decreasing $\alpha$ decreases the radio emission
in Sgr A*.  This is because, near the central object, $\rho \propto
\dot m$, and is only a weak function of $\alpha$.  The density on the
outside, however, scales as $\rho \propto \dot m/\alpha$.  If we fix
the X-ray flux, $\mo$ has to decrease as $\alpha$ decreases in order
to keep $\rho$ the same on the outside and thereby produce the same
level of bremsstrahlung radiation.  This causes a decrease in the
density in the interior of the flow and thus a decrease in the
synchrotron emission in the radio (Figure 5a).  Small values of
$\alpha$ therefore add to the decrease in synchrotron emission that is
associated with a strong wind.

Finally, Figure 6 shows models of Sgr A* with no winds ($p = 0$) for
several $\alpha \sim \beta^{-1}$ (Fig. 6a) and for several $\delta$
(Fig.  6b).

\subsection{Comparison with Observations}

Figures 4-6 show the observational data on Sgr A$^*$.  The source has
been reliably detected only in the radio and mm bands, where there is
a good spectrum available (see Narayan et al. 1998a for original
references to the data).  It has been convincingly demonstrated that
there is a break in the radio spectrum at around 50--100 GHz, so that
the source apparently has two components, one which produces the
emission below the break and the other above (Serabyn et al. 1997,
Falcke et al. 1998).  The latter component, which cuts off steeply
somewhere between $10^{12}$ and $10^{13}$ Hz, has been fitted with the
ADAF model (Narayan et al. 1998a).  The model does not, however, fit
the low frequency radio emission.  This emission may be from an
outflow (e.g.  Falcke 1996), or, as in the model of Mahadevan (1998),
may be due to non-thermal electrons (in Mahadevan's model, these,
along with positrons, are created by the decay of charged pions
created in proton-proton collisions).  In the following we consider a
model to be satisfactory if it fits the high frequency radio data.

In the infrared, Menten et al. (1997) obtained a conservative 2.2
micron flux limit of 9 mJy, after accounting for extinction.  The
source may, however, be variable, since in later epochs Genzel et al.
(1997) observed a $K\sim15$ source at the location of Sgr A$^*$; this
corresponds to $F_\nu \approx 13$ mJy (Andreas Eckart, private
communication).  If verified, this would suggest that the infrared
flux varies around a mean value of order a few mJy.  This is a
potentially stringent constraint on theoretical models.  We, however,
adopt a more conservative approach and treat the IR data as an upper
limit.  The implications of an IR detection are discussed in \S6.

Although we fit our models to the ROSAT X-ray observations of the
galactic center, they too should be treated as an upper limit because
of ROSAT's poor angular resolution ($\approx 20$'') and the presence
of diffuse emission at the Galactic Center.  This is again the
conservative approach, since a decrease in the X-ray flux would
necessitate a decrease in the importance of mass loss; see \S6.

Vargas et al. (1998) have recently provided new SIGMA upper limits on
hard X-ray emission from the Galactic Center: between $40-75$ keV the
luminosity is $\lsim 1.4 \times 10^{35}$ ergs s$^{-1}$ while between
$75-150$ keV it is $\lsim 2.0 \times 10^{35}$ ergs s$^{-1}$.  We have
converted these to limits in $\nu L_\nu$ by assuming that the spectrum
is flat in $L_\nu$, as would be appropriate for a no-wind
bremsstrahlung spectrum.

The solid line in Figure 4a (and the dotted line in Figure 6) shows
our standard, no-wind ($p=0$), model, with $\beta = 10$, $\alpha =
0.1$, and $\delta = 0.01$.  Figure 6 shows no-wind models for a number
of other microphysics parameters.  All of the no-wind models are in
reasonable agreement with the data.  In particular, they explain the
mm fluxes fairly well as synchrotron emission, and produce Compton
emission in the infrared roughly consistent with the Menten et
al. (1997) limit.  Relatively lower $\delta$, larger $\beta$, and
smaller $\alpha$ are favored if the IR limit is taken to be stringent;
if, however, the Genzel et al. (1997) observations are interpreted as
a detection, the opposite is true --- larger $\delta$ and/or smaller
$\beta$ are favored.  In addition, the small $\beta$, large $\delta$
models tend to slightly overproduce the synchrotron emission at $\sim
10^{12}$ GHz.

Note that these conclusions are somewhat different from those of
Narayan et al. (1998a), who advocated strict equipartition ($\beta =
1$).  As discussed in \S2, this is due to our use of a monatomic ideal
gas adiabatic index in the electron energy equation.  For small
$\delta$, the electrons in Sgr A* are nearly adiabatic; since our
adiabatic index is larger than that of Narayan et al. the electrons
are hotter in our models.  This accounts for the increased synchrotron
emission and the need for weaker fields (larger $\beta$) for a fixed
radio flux.  To obtain a radio flux comparable to Narayan et al's
$\beta = 1$ model, we require $\beta \approx 30$ for $p = 0$, or else
$p \approx 0.2$.  In fact, our no-wind, low $\delta$, models of Sgr A*
are rather similar to those of Manmoto et al. (1997), who noted that
smaller $\alpha$ were favored if the IR limit in Sgr A* is taken to be
stringent.  This is because our treatment of the electron energy
equation is similar to Manmoto et al.'s.  They took the electron
adiabatic index to be $\gamma_e = 5/3$, which is correct in not
including a magnetic contribution.\footnote{It is incorrect, however,
in neglecting the change to $\gamma_e \approx 4/3$ for $r \lsim 10^2$
when the electrons become semi-relativistic.}

What about large $p$, dynamically important winds?  Such winds
decrease the density and electron temperature in the interior of the
flow, thereby severely suppressing the synchrotron and Compton
emission (Figure 4a).  Requiring wind models to produce the observed
$10^{11}- 10^{12}$ Hz emission imposes the following strong
constraints on the parameters.

For small $\delta$, we require $p \lsim 0.2$ if we allow $\beta \sim
1$, $\alpha \sim 0.3$.  If, for theoretical reasons, we were to favor
larger $\beta \sim 10-100$, the constraint is even stronger.  For the
value of $\ro = 10^4$ used in our models, this corresponds to at least
15 \% of the mass supplied at large radii reaching the central object.
As in V404 Cyg, the strongest degeneracy in the problem is with
$\delta$.  For $\delta \gsim 0.3$, large $p$ models of Sgr A* are in
good agreement with the data (Figure 5b).

All ADAF models of Sgr A* in the literature have $\mo \sim 10^{-4}$.
This is at the lower end of the values considered plausible from Bondi
capture of stellar winds in the Galactic center, and may be $\sim
10-100$ times smaller than favored values (Coker \& Melia 1997).  It
is interesting to see that winds do not alter this conclusion (see
Table 2).  Neither wind nor non-wind models can have $\mo$ much
greater than $\sim 10^{-4}$ because, if they did, the bremsstrahlung
emission would yield an X-ray luminosity well above the observed
limits.  Since the bremsstrahlung emission at $\sim 1$ keV is from the
largest radii in the accretion flow, this conclusion is independent of
the strength of winds in the system.

In this context, it is important to note that, although $p = 0$, large
$\delta$ models produce spectra reasonably consistent with the
observations (Fig. 6b), they require small accretion rates, $\mo \sim
10^{-5}$ (Table 2).  This argues against them as viable models.

\section{Nuclei of Nearby Ellipticals}

D98 recently measured high frequency radio fluxes from the nuclei of
several nearby giant elliptical galaxies.  These galactic nuclei are
known to be unusually dim in X-rays compared to the accretion rates
inferred from Bondi capture (Fabian \& Canizares 1988).  Fabian \&
Rees (1995) explained the low X-ray luminosities by invoking accretion
via ADAFs.  D98 found, however, that the predicted radio emission,
based on the ADAF model (for $\beta=1$), exceeded their measured
fluxes by $2-3$ orders of magnitude.  They suggested several
explanations for this large discrepancy, including the presence of
strong winds or highly subthermal magnetic fields.

If, as we are inclined to believe is the case, Sgr A* is simply a
scaled version of the systems observed by D98, why is the predicted
emission in Sgr A* roughly in accord with the observations while that
in the nearby ellipticals is so discrepant?  We might expect both
theoretical predictions to be wrong, or both to be right, if the same
physics operates in each system.  We see two potential answers to this
question.

One possibility lies in the X-ray constraints in Sgr A* versus those
in D98's sample.  In Eddington units, i.e, scaled with respect to the
mass of the black hole, the X-ray detection/upper limit in Sgr A* is
$\sim 2.5$ orders of magnitude below the upper limits in D98's sample.
This means that we have a significantly stronger constraint on the
accretion rate in Sgr A$^*$.  If the X-ray luminosities (in Eddington
units) of the ellipticals were as low as in Sgr A$^*$, then models
similar to those that work for Sgr A$^*$ would work for the
ellipticals as well.  It would mean, however, that D98's estimate of
$\mo$ is too large, by a factor $\sim 30$ (see below).

The other possibility is that the high frequency ($>10^{11}$ Hz) radio
observations of Sgr A*, which the ADAF model fits reasonably well, do
not probe the accretion flow at all.  If the high resolution VLBI
observations at 86 GHz represent the true synchrotron emission from
the ADAF in Sgr A*, and the higher frequency radio emission is from a
completely different source, then typical no-wind (e.g. $p = 0$,
$\beta \sim 1$) models would overpredict the synchrotron luminosity by
$\sim 3$ orders of magnitude, just as D98 found for the ellipticals.

To investigate these issues further, Figure 7 shows a series of models
of NGC 4649, which D98 consider to be the most convincing member of
their sample.  The data are taken from their Table 5.  We take $m = 8
\times 10^9$ (slightly higher than D98's $4 \times 10^9$ because we
find this mass fits the location of the radio peak better), $\ro =
10^4$, and assume a distance of $15.8$ Mpc.  All calculations were
done with $\alpha = 0.1$ and $\beta = 10$.  Table 3 shows the
parameters for the models.

The solid line in Figure 7a corresponds to our ``standard'' ADAF
model: $p = 0$, $\delta = 0.01$, and $\mo = 10^{-3}$.  The latter
value corresponds to the Bondi mass accretion rate estimated by D98.
In agreement with D98, we find that, at this accretion rate, the model
overpredicts the radio emission by $\sim$ 3 orders of magnitude.  To
make matters worse, our model is also in violation of the X-ray upper
limit, in contrast to D98, whose ``standard'' ADAF model just
satisfies the upper limits.  The difference is primarily because our
electrons are hotter --- D98 used Esin et al's (1998) electron
adiabatic index.

We have varied $p$ and $\mo$ in our models to judge their sensitivity
to these parameters.\footnote{Initially, we found important
quantitative differences between our models with varying $p$ and $\mo$
and the models in the original version of D98's paper on astro-ph.  We
have determined, however, that this was due to the fact that they did
not use the electron temperature profile appropriate for the given
$\mo$ and $p$ (Di Matteo, private communication).  In particular, they
originally required $p = 1$ and $\ro = 300$ to fit the radio flux at
$\mo = 10^{-3}$, while their new calculations give $p \approx 0.8$ and
$\ro \approx 80$ (since they take the inner radius of the flow to be
$r = 3$, this corresponds to $\approx 7 \%$ of the incoming mass
accreted, comparable to our value of $\approx 10 \%$).  In addition,
at $p = 0$, they originally required $\mo = 10^{-6}$ to fit the radio
flux, while they now require $\mo \approx 10^{-5}$, again in
reasonable agreement with our value of $\mo \approx 10^{-4.5}$.}  The
dotted line in Figure 7a is a model with $p = 0.25$ and $\delta =
10^{-2}$.  This is roughly the $p$ we need to account for the observed
radio flux at low $\delta$ (note that this model is also in agreement
with the X-ray upper limit).

In Figure 7b we show several models of NGC 4649 for $\mo = 10^{-4.5}$.
This accretion rate is $\sim 30$ times smaller than the value D98
infer from Bondi capture.  The solid line shows a standard no-wind
model: $p = 0$ and $\delta = 0.01$.  This model is in reasonable
agreement with the radio flux.  If $T_e$ and $\beta$ are fixed,
equation (\ref{lsynch}) shows that the peak synchrotron luminosity
scales like $\nu L_\nu \propto \dot m^{3/2}$.  Thus, to decrease $\nu
L_\nu$ by a factor of $10^3$, as required by the observations, $\mo$
must decrease by $\sim 100$.  In fact, due to other factors, the
required decrease is even less, $\sim 30$.

The above argument requires that $T_e(r)$ should be roughly the same
for $\mo = 10^{-3}$ and for $\mo = 10^{-4.5}$.  This is confirmed by
the numerical results shown in Table 3, but it can also be understood
simply by noting that in both models the electrons adiabatically
compress as the gas flows in.  To see this, it is sufficient to
estimate the PdV energy gained per unit time by the electrons in a
spherical shell of radius $R$ and thickness $dR \sim R$ as they
accrete onto the central object (cf eq. [\ref{ee}]), \beq q_c \approx
k T_e v {d n_e \over dR} 4 \pi R^3 \approx {m_e \over m_p} \theta_e(r)
\dot M(r) c^2 \approx 10^{43} \left({\theta_e \over 1}\right)
\left({\dot m \over 10^{-3}}\right)\left({m \over 8 \times
10^9}\right) {\rm ergs \ s^{-1}}.
\label{comp} 
\eeq Our most luminous model (solid line, $\mo = 10^{-3}$; Figure 7a)
has $\nu L_\nu \approx 10^{41.5} {\rm ergs \ s^{-1}}$ at the
synchrotron peak (and a bolometric luminosity of $\approx 10^{42} {\rm
ergs \ s^{-1}}$), which is $\ll q_c$.  For lower $\mo$, the ratio of
$\nu_s L_{\nu,s}$ to $q_c$ is even smaller.  Therefore, the electrons
in all of our low $\delta$ models are nearly adiabatic, and thus
$T_e(r)$ is essentially unchanged as $\mo$ decreases from $10^{-3}$ to
$10^{-4.5}$.

The short dashed lines in Figure 7 show $p = 0.25$, $\delta = 0.3$
models for $\mo = 10^{-3}$ (Figure 7a) and $\mo = 10^{-4.5}$ (Figure
7b).  As suggested by the previous results on V404 Cyg and Sgr A$^*$,
these models are comparable to the $p = 0$, $\delta = 0.01$ models.
In particular, for $\mo = 10^{-4.5}$, the wind model gives reasonably
good agreement with D98's radio data, while for $\mo = 10^{-3}$ it is
in disagreement.
 
The results of Figure 7 thus lead to two scenarios for understanding
NGC 4649, depending on which value of $\mo$ we take, $10^{-4.5}$ or
$10^{-3}$.  (There is, of course, a range of intermediate scenarios if
we take intermediate values of $\mo$.)

If $\mo \approx 10^{-4.5}$, then we require $0 \lsim p \lsim 0.25$ for
$0 \lsim \delta \lsim 0.3$.  As we saw for V404 Cyg and Sgr A$^*$,
increasing $p$ requires a corresponding increase in $\delta$, though
the precise mapping between the two parameters in the case of NGC 4649
is slightly different.  As in V404 Cyg and Sgr A$^*$, strong wind, low
$\delta$ models are ruled out as they cannot explain the radio data
(dotted line; Figure 7b).

If, on the other hand, $\mo \approx 10^{-3}$, as proposed by D98, then
we require $0.25 \lsim p \lsim 0.55$ for $0 \lsim \delta \lsim 0.3$.
Low $\delta$, low $p$, is ruled out by the observed radio flux (Figure
7a), which is a different result from that obtained in V404 Cyg and
Sgr A$^*$.  In addition, the region of $p-\delta$ space available for
NGC 4649 at $\mo = 10^{-3}$, if applied to our models of V404 Cyg and
Sgr A$^*$, is somewhat uncomfortable.  For example, at $p \approx
0.25$ and low $\delta$ (which gives an acceptable fit in NGC 4649 if
$\mo \approx 10^{-3}$), the predicted X-ray spectral index in V404 Cyg
is only marginally compatible with the 2 $\sigma$ ASCA measurements
(Figure 1a).  Similarly, the radio luminosity of Sgr A* for $p = 0.25$
and $\delta = 0.01$ is $1-2$ orders of magnitudes below the peak
observed luminosity.  One might therefore have to abandon the claim
that the $10^{11}- 10^{12}$ Hz emission in Sgr A$^*$ is synchrotron
emission from the ADAF.

If we believe that Sgr A*, V404 Cyg, and NGC 4649 are simply scaled
versions of each other (in $m$ and $\mo$, and perhaps somewhat in $p$,
$\delta$, $\beta$, $\alpha$, and $\ro$), the above considerations are
suggestive, if only weakly, of an $\mo \sim 10^{-4.5}$ rather than
$10^{-3}$ in NGC 4649.  This conclusion is independent of the
importance of winds.  

\section{Key Future Observations}

There are two main conclusions from the previous sections: (i) If the
electron heating parameter $\delta$ is small, current observations
rule out ADAF models with moderate winds (say $p\gsim0.25$ as an
average for V404 Cyg and Sgr A*).  (ii) If $\delta$ is allowed to have
large values --- given the uncertain role of magnetic reconnection
there is no strong theoretical argument against this --- current
observations provide no information on the importance of winds in
ADAFs; large $\delta$, strong wind models are in as good agreement
with the data as low $\delta$, weak wind models.

Figure 8 shows the $p/\delta$ degeneracy explicitly for V404 Cyg
(Fig. 8a) and Sgr A* (Fig. 8b).  We see that the two $p = 0.4, ~\delta
\approx 0.3$ models are very similar to the $p = 0, ~\delta = 0.01$
models.  Indeed, there is a family of intermediate solutions with
values of $p$ and $\delta$ in between these two extremes.  Note,
however, that the very large $p = 0.8, ~\delta = 0.75$ models shown in
Figure 8 differ more noticeably.  For such large $p$, the electron
temperatures needed to make the X-ray spectrum of V404 Cyg Compton
dominated, rather than bremsstrahlung dominated, are so large that the
Compton peak moves well into the X-ray band.  This is discussed
further in the next subsection.  In the case of Sgr A*, for $p \sim
1$, $\ro \sim 10^4$ and $\mo \sim 10^{-4}$, the inner mass accretion
rate is $\dot m_{in} \sim 10^{-8}$.  The density in the interior is
then so low that the synchrotron emission is no longer highly
self-absorbed; this accounts for the substantially broader synchrotron
peak.

Leaving aside the $p=0.8$ models, we conclude that there is a
degeneracy between $p$ and $\delta$ such that any model in the range
$0<p\lsim0.5$ and $0<\delta\lsim0.4$ is viable, so long as $p$ and
$\delta$ are chosen in some proportionate manner.  Current
observations are insufficient to break this degeneracy and additional
observational tests are clearly needed to improve the situation.
Below we suggest a number of such tests, some of which will be
feasible in the near future (e.g., AXAF; \S6.3-\S6.5), while others
will require a more concerted observational effort (e.g., \S6.1 \&
\S6.2).

\subsection{Position of the Compton Peak}

Generically, strong wind models have a lower density and optical depth
than weak wind models.  If these models are to fit the synchrotron and
X-ray flux, and have a Compton-dominated X-ray spectrum (as required
in SXTs for instance), they must have a larger $T_e$.  The larger
$T_e$ is necessary to boost the synchrotron luminosity, and also to
reproduce the required $\alpha_c$ (cf eq. [\ref{compt}]), despite the
smaller optical depth.  A larger $T_e$ implies a larger amplification
factor $A$.  Thus, the ``distance'' between the synchrotron and
Compton peaks in large $\delta$ wind dominated systems will generally
be larger by a factor of a few than in weak wind systems.  This is
seen explicitly in Figure 8a.  Note in particular the $p = 0.8$,
$\delta = 0.75$ model, where $A$ is so large that the synchrotron peak
has moved substantially to the left and at the same time the peak of
the Compton emission is well into the X-ray band.  This is, however,
not a unique property of wind models, but rather is characteristic of
any model with a very high $T_e$, as can be seen by the $\delta =
0.3$, $p = 0$ model in Figure 3b.

In principle, the Compton A parameter could be measured in SXTs, with
observations in the optical and soft X-ray ($\sim 0.1$ keV) bands.
The strong Galactic absorption below a keV, however, makes direct
detection of the Compton peak problematic, except in very high
$\delta$ models.  A more encouraging possibility is that the peak's
position could be inferred by detection of curvature at $\sim 1$ keV.
This should be feasible with AXAF or XMM (\S 6.4).

\subsection{Shape of the Bremsstrahlung Spectrum}

Detailed measurements of the bremsstrahlung spectrum of an ADAF system
can explicitly probe the density profile and outer radius of the
accretion flow.  This constitutes one of the most direct tests for the
presence of winds.

As discussed in \S2, at photon energies $\gsim$ the minimum electron
thermal energy in the flow, which is a function of the outer radius,
bremsstrahlung should give rise to a $\nu L_\nu \propto \nu^{1/2 -
2p/\epsilon}$ spectrum (where $\epsilon \approx 1$; see \S2.2).  This
behavior is clearly seen in Figures 1a (V404 Cyg), 4a (Sgr A*), and,
in particular, Figure 8b (Sgr A*), where the bremsstrahlung emission
cuts off strongly at $\gsim 10$ keV for large $p$.  More importantly,
the details of the cutoff, e.g., the slope at $\sim 10$ keV, are a
strong function of $p$, thus providing the opportunity to study winds
directly through their effect on the density profile of the flow.
While energies $\gsim 10$ keV are somewhat high for observations of
quiescent systems with current X-ray detectors, they may still be
observationally accessible.  Sgr A* is an excellent source in which to
apply this technique since a bremsstrahlung-dominated X-ray spectrum
is expected above a few keV for a wide range of the microphysical
parameters.  The SXTs are probably less useful, since the Compton peak
is generally more important.

One potential complication is that, if winds only become important
well inside the outer boundary of the flow (for which, at least within
the context of BB98's proposal, we see no obvious theoretical reason),
the bremsstrahlung emitting region will be unaffected by the wind.
Non-detection of a strong $\sim 10$ keV bremsstrahlung cutoff
therefore would not rule out winds, although it would place
interesting constraints on the radius at which mass loss becomes
important.

\subsection{Observations of SXTs in Quiescence}

Measurements of the X-ray spectral index in SXTs by AXAF and, in
particular, XMM (with its larger collecting area) will clearly be
important.  Bremsstrahlung predicts a photon spectral index of $\Gamma
\sim 1$, while Comptonization predicts a less hard spectrum.

ASCA observations rule out $\Gamma \sim 1$ in V404 Cyg; therefore,
models in which Comptonization dominates are favored (Narayan et
al. 1997a).  Confirmation of this in additional systems would be of
considerable interest.  Note that detection of a bremsstrahlung
spectrum would all but rule out weak wind ($p \approx 0$) models of
quiescent SXTs.  As Figure 3 shows, for no combination of $\alpha,
\beta$ and $\delta$ can such a spectrum be produced.

When Comptonization dominates, the $0.5-10$ keV X-ray band is often
the energy range where the first and second Compton bumps meet.
Therefore, most models predict significant curvature, i.e., an energy
dependent spectral index, within the band.  The spectrum would be
softer at lower photon energies and harder at higher energies.
Detection of curvature of this sign would be an important confirmation
of the Compton origin of the X-ray emission.  Curvature of the
opposite sign will be seen only if $\delta$, and therefore the Compton
$A$-parameter, is so large that the Compton peak moves into the middle
of the X-ray band (\S6.1).

It is equally important to measure the optical/UV spectrum of
quiescent SXTs.  An unambiguous determination that the temperature of
the optical radiation is above the maximum temperature $\sim5000$ K of
the outer thin disk (Lasota et al. 1996), would immediately suggest
that the optical emission is synchrotron emission from the ADAF.  This
information would be very useful, since only a subset of models will
be able to fit both the observed synchrotron power and the position of
the peak.  

Although significantly tighter observational constraints on the
energy-dependent spectral index in X-rays and the synchrotron peak in
the visible/UV would narrow the available theoretical parameter space,
it will be difficult to infer unique model parameters from these
observations.  There is simply too much degeneracy in the models.  In
particular, strong wind, large $\delta$ models of SXTs can readily
produce X-ray and optical behavior similar (though not identical) to
that of weak wind models (see Figure 8a).  The variations introduced
by $\alpha$ and $\beta$ only complicate things further (Figure 3).

\subsection{Observations of Sgr A*}

\subsubsection{X-ray Observations}

Observations of Sgr A* may be particularly helpful in discriminating
among theoretical models.  An AXAF/XMM detection of Sgr A* will help
in two ways.  First, a significantly higher resolution detection will
determine whether the ROSAT observation was an upper limit.  If
significantly so, $\mo$ will need to decrease in order to fit the
reduced X-ray flux; the synchrotron emission in the radio will
decrease as well (all other parameters being held fixed).  In the case
of no winds ($p = 0$), this would argue for stronger magnetic fields
and/or larger $\delta$.  In wind scenarios, the amount of mass loss in
the flow would be further constrained and/or the models would be
pushed towards larger $\delta \sim 1$.

As Figures 4--5 and Figure 8b show, all of the models of Sgr A* with
noticeable winds have a bremsstrahlung dominated X-ray spectrum with
$\Gamma \sim 1$.  This is true even for the large $\delta$ models
which agree with the radio observations (Fig. 8b).  Confirmation of
this prediction would be of considerable interest.  It would not,
however, rule out weak wind scenarios, although strictly $p = 0$
models would be constrained to having small $\alpha$ and $\delta$ and
large $\beta$ (see Figure 6).\footnote{Even at $p = 0.2$ (a relatively
weak wind in the scheme of things) the X-ray band would be
bremsstrahlung dominated for a wide range of microphysical parameters,
eliminating these constraints.}  On the other hand, an observed
spectral index deviating significantly from bremsstrahlung, e.g.,
$\Gamma \gsim 2$, would strongly constrain mass loss via winds,
arguing for $p \lsim 0.2$ and/or larger $\delta$.

\subsubsection{Infrared Observations}

Additional constraints on the accretion in Sgr A* may come from
infrared observations.  Menten et al. (1997) obtained strong upper
limits on the 2.2 $\mu m$ flux from Sgr A* at an angular resolution of
$0.15''$, while Genzel et al. (1997) reported a possible infrared
detection of Sgr A* at a level above the Menten et al. limit,
indicating possible variability in the source.

As seen in Figures 4-6 \& 8b, the 2.2 micron K band corresponds to the
location of the first Compton peak (if present) in ADAF models.  If
Genzel et al.'s detection is confirmed, and interpreted as the Compton
peak, it would argue against significant mass loss due to winds in Sgr
A*.  As seen in Figures 4, 5, \& 8b, large $p$ models have difficulty,
even at large $\delta$, accounting for a Compton luminosity at the
level of the Menten et al. (1997) upper limit ($p\lsim 0.2$ and large
$\delta \sim 1$ is tenable).  This is because the Compton power
decreases more rapidly with $p$ than the synchrotron power.
Increasing $\delta$ to bring the synchrotron power back into agreement
still leaves the Compton power smaller than in no-wind models (see
Fig. 8b).

Equally interesting constraints on models will arise if the infrared
limits decrease substantially.  This would argue for weak magnetic
fields and/or small $\delta$ in no wind scenarios (see Figure 6) or
for the presence of a reasonable wind.  If the IR limits decrease
substantially, it may be difficult for theoretical models to reconcile
the absence of a Compton peak (the source of IR emission in the
models) with the interpretation of the $\sim 10^{12} $ Hz emission as
synchrotron emission from the flow.

An alternative to the Compton interpretation of the infrared flux is
that it is due to synchrotron emission from non-thermal electrons.
Mahadevan (1998) has described a specific model of this kind in which
the infrared flux is from electron/positron pairs created by the decay
of charged pions in the accretion flow.

\subsubsection{Gamma-ray Observations}

Mahadevan et al. (1997) showed that gamma-ray emission, due to the
decay of neutral pions created in proton-proton collisions, may be
detectable from an ADAF in Sgr A*.  They argued that it may account
for the EGRET source 2EG J1746-2852, although current theoretical
predictions suggest that the ADAF contribution to these observations
may be small (Narayan et al. 1998a, Mahadevan 1998; see Markoff et
al. 1997 for an alternative discussion of gamma-ray emission from Sgr
A*).  Since the gamma-ray luminosity is $\propto \rho^2$, the
confirmed detection of gamma-rays from Sgr A$^*$ (i.e., a detection
with reasonable angular resolution which shows variability) would
directly constrain the density of the plasma and thus the strength of
winds.  GLAST may provide such observations.

A complication in this analysis is the unknown, and almost certainly
non-thermal (Mahadevan \& Quataert 1997, Blackman 1998), proton
distribution function.  If there is a significant population of
relativistic protons at all radii in the flow, the emission will
contain important contributions from large radii (Mahadevan et
al. 1997), and will thus be a complicated convolution of the density
profile of the flow and the (radially varying) proton distribution
function.  If, on the other hand, relativistic protons are only
present in the interior of the flow, e.g., if the protons are heated
by incompressible turbulence (Gruzinov \& Quataert 1998), gamma-ray
observations would impose particularly strong constraints on the flow
properties near $r \sim 1$.  Specifically, they will provide a lower
limit on the mass accretion rate near the black hole, which, when
combined with the estimate of $\mo$ obtained from measurements of the
bremsstrahlung emission, would strongly constrain the wind parameter
$p$.

\subsubsection{Measurements of Radio Brightness Temperatures}

VLBI observations of Sgr A* at 43 GHz and 86 GHz indicate brightness
temperatures in excess of $10^{10}$ K (Backer et al. 1993, Rogers et
al. 1994).  This was a problem for Narayan et al's (1998a) no wind
model since their electron temperature was everywhere below $10^{10}$
K.  With the revised electron adiabatic index used in this paper
(\S2), we find $T_e \gsim 10^{10}$K for $r \lsim 10$, so this is
less of a concern.

Small $\delta$, large $p$ models give $T_e \lsim 5 \times 10^9$ K at
all radii in the flow (see Table 2), incompatible with the
observations.  This is because, as noted in \S2.3, large $p$ implies
less compression and hence lower $T_e$.  These low values of $T_e$ are
an independent argument against high $p$, low $\delta$ models.

For $\delta \gsim 0.1$, turbulent heating of the electrons is
sufficiently strong that $T_e$ is again $\gsim 10^{10}$K, compatible
with the observations.  For a fixed radio flux, large $p$ models must
have higher electron temperatures to compensate for the lower flow
densities; the difference is, however, only a factor of a few.
Nonetheless, a careful comparison of theoretical and observed
temperature profiles may help constrain the models.  If brightness
temperatures $\gsim 3 \times 10^{10}$K are observed, they would argue
for larger $\delta$, in either wind or non-wind scenarios, or for
non-thermal electrons.

\subsection{Observations of the Nuclei of Nearby Ellipticals}

As is clear from Figure 7, one way to resolve the degeneracy between
$\mo \approx 10^{-3}$ and $\mo \approx 10^{-4.5}$ models of NGC 4649,
and the associated $p$/$\delta$ degeneracy, is through better X-ray
limits in this and other nearby ellipticals.  The $\mo \approx
10^{-3}$ scenario predicts an X-ray flux comparable to the current
limits, so AXAF or XMM should be able to detect these systems.  On the
other hand, the $\mo \approx 10^{-4.5}$ scenario predicts X-ray fluxes
$\sim 3$ orders of magnitude beneath current limits.

D98's observations in the radio and mm bands with the VLA and SCUBA
emphasize the usefulness of such observations for testing ADAF models.
They allow both the synchrotron power and the position of the peak to
be measured.  Additional observations of low luminosity nuclei,
including LINERs (Lasota et al. 1996a), will further constrain
theoretical models.

\section{Summary}

The goal of this paper has been to explore spectral models of ADAFs
with winds/mass loss, by applying the models to the soft X-ray
transient V404 Cyg, the Galactic Center source Sgr A*, and the nucleus
of the nearby elliptical galaxy NGC 4649.  The first two of these
systems are explained fairly well by the ADAF model without any winds.
However, recent theoretical arguments (BB98; see also Narayan \& Yi
1994, 1995), as well as observations of NGC 4649 (D98), suggest that
mass loss via winds may be important (even dynamically crucial) in
sub-Eddington, radiatively inefficient, accretion flows.  We have
therefore investigated under what conditions ADAF models with mass
loss might account for the observations, and the extent to which
observations can distinguish between the various proposals for the
physics of the accretion flow.  A fundamental assumption of our
analysis is that, in spite of the possibility of substantial mass
loss, the observed radiation is due only to the matter that accretes
onto the central object, with no contribution from the outflow.

In assessing the importance of mass loss, considerable care must be
taken in how one treats the microphysics of the accretion flow; this
is parameterized by: $\alpha$, the viscosity parameter, $\delta$, the
fraction of the turbulent energy which heats the electrons, and
$\beta$, the ratio of the gas to the magnetic pressure.  As we have
shown in this paper, mass loss from the accretion flow has a dramatic
effect on theoretically predicted spectra.  If we were to restrict
ourselves to the microphysics parameters considered in previous
treatments of ADAFs ($\alpha \sim 0.3$, $\beta \sim 1$, and $\delta
\sim 10^{-3}$), significant mass loss would be incompatible with
observations.  Such a restriction would not, however, correctly
reflect the uncertainty in the microphysics of the flow.  We have
therefore varied $\alpha$, $\beta$ and $\delta$ over a large range
($\alpha \ \epsilon \ [0.03, 0.3]$, $\beta \ \epsilon \ [1, 100]$, and
$\delta \ \epsilon \ [0, 1]$), which we believe generously encompasses
the theoretical uncertainties.  Despite this large parameter space,
firm and interesting constraints on theoretical models can still be
drawn.

Spectral models of ADAFs without mass loss provide a reasonable
description of the observations of a number of low luminosity black
hole systems.  In fact, it is this success which has led to the recent
wide interest in ADAFs.  From our study of spectral models that
include mass loss/winds, we reach two principal conclusions.  First,
if the turbulent heating of the electrons is weak ($\delta
\lsim0.01$), then winds must also be relatively weak ($p\lsim0.25$);
specifically, the mass accreted by the central black holes in V404 Cyg
and Sgr A$^*$ must be at least $\sim 10 \%$ of the mass supplied at
large radii (for any $\alpha$ and $\beta$).  Second, for larger values
of $\delta$, current observations do not readily assess the importance
of winds in ADAFs.  Strong wind ($p=0.4$), large $\delta$ ($\sim0.3$)
models of V404 Cyg and Sgr A* are in as good agreement with the data
as weak wind ($p\sim0$), low $\delta$ ($\lsim0.01$) models; indeed,
there is a family of acceptable models in which $p$ and $\delta$ are
assigned intermediate values in a proportionate manner.

In \S6, we have proposed a number of observational tests which should
help to resolve the $p/\delta$ degeneracy.  Radio, infrared,
optical/UV, X-ray and gamma-ray observations of quiescent soft X-ray
transients, the Galactic Center, and the nuclei of other galaxies
provide powerful and complementary information on the various emission
processes in the accretion flow and can potentially pin down the
importance of winds in ADAFs.

D98's observations of radio emission in nearby ellipticals such as NGC
4649 seem to suggest that mass loss may be important; confirmation of
this interpretation of their observations, however, requires
substantially better X-ray observations than are presently available.
Current X-ray upper limits and radio observations alone are compatible
with weak (or no) mass loss at accretion rates $\sim 30$ times smaller
than the typical Bondi value inferred by D98 (\S5).  X-ray detection
of these systems near current limits would favor high accretion rates
and substantial mass loss while noticeably stronger upper limits would
favor low accretion rates and weaker mass loss (\S6.5; Figure 7).

In this context, it is interesting to note that Sgr A* itself may have
a discrepancy between the accretion rate favored by ADAF models and
that favored by hydrodynamical simulations of Bondi capture.  All
non-wind ADAF models in the literature have $\mo \sim 10^{-4}$, while
Bondi capture estimates are often $10-100$ times larger (e.g., Coker
\& Melia 1997).  Winds do not alleviate this discrepancy (see Table
2); neither strong wind nor non-wind ADAF models of Sgr A* can have
$\mo$ much greater than $\sim 10^{-4}$ because, if they did, the
bremsstrahlung emission would yield an X-ray luminosity well above the
observed limits.

\subsection{Radiative Efficiency}

Previous discussions of ADAFs in the literature have emphasized the
low radiative efficiency of these accretion flows and the connection
between the low efficiency and the presence of an event horizon in the
central object (Narayan et al. 1997ab, Narayan et al. 1998a, Menou,
Quataert \& Narayan 1998).  How is this modified when there is a wind?

Tables 1-3 give the radiative efficiencies of our models, defined both
with respect to the accretion rate at the outer edge of the flow
($\eta_o \equiv L/\dot M_{\rm out} c^2$) and the accretion rate at the
horizon ($\eta_i \equiv L/\dot M_{\rm in} c^2$).  The former is
perhaps observationally accessible, while the latter is of more
physical interest from the point of view of inferring the presence of
a horizon.  In the absence of winds, $\eta_o = \eta_i \equiv \eta$.

Qualitatively, a viable spectral model with strong winds weakens the
argument for the presence of an event horizon.  This is because, in
the presence of a strong wind, the central object accretes less mass,
but if the microphysical parameters can be adjusted so that the
accretion flow produces the same luminosity, then $\eta_i$ can be
noticeably larger.  What remains to be seen is, quantitatively, how
$\eta_i$ varies with $p$ and the microphysical parameters.

The models of most interest are those in Figure 8, the sequence of
roughly similar $p/\delta$ models of V404 Cyg and Sgr A$^*$.  The
standard no wind models of V404 Cyg and Sgr A$^*$ have
$\eta\approx10^{-3}$ and $2.6 \times 10^{-5}$, respectively, which
makes both sources highly inefficient radiators.  As we increase $p$,
and correspondingly increase $\delta$ to keep a roughly similar
luminosity, $\eta_i$ increases.  For the intermediate values of $p =
0.4$ and $\delta \sim 0.3$, $\eta_i$ has increased by nearly an order
of magnitude for both systems.  The radiative efficiency is, however,
still $\ll 1$, particularly for Sgr A*.

For the larger values of $p = 0.8$ and $\delta = 0.75$, $\eta_i
\approx 0.1$ for V404 Cyg and $\eta_i \approx 0.01$ for Sgr A*.  The
model of V404 Cyg appears to be incompatible with the observations.
The synchrotron peak in the optical has shifted to much lower
frequencies in this model and we believe that this discrepancy is
serious and hard to overcome.  Nevertheless, if we are willing to
allow models with this level of deviation from the data (there are,
after all, uncertainties in our modeling techniques in this extremal
range of parameter space), then it means that there are viable ADAF
models of V404 Cyg in which the luminosity of the accretion flow is
comparable to the rest mass energy accreted by the central object.
This weakens the argument for an event horizon in this source.

In the case of Sgr A$^*$, the $p = 0.8$, $\delta = 0.75$ model is in
moderate agreement with the data (given some allowance for modeling
uncertainties), but it is still a relatively inefficient radiator
($\eta_i \approx 0.01$).  The difference between Sgr A* and V404 Cyg
in this context is, of course, that $\mo$ is quite a bit smaller in
Sgr A* (because of the constraints imposed by X-ray observations), so
that it naturally has a lower radiative efficiency if all other model
parameters are the same.  In fact, we find that all of our models of
Sgr A* with $\mo \sim 10^{-4}$, which fit the observations reasonably
well, have $\eta_i \lsim 0.01$.  At small $p$, this is because only
small values of $\delta$ are allowed if $\mo \sim 10^{-4}$ (see Table
2).  At large $p$, the density in the interior of the flow is so low
that, even if $\delta \equiv 1$ or the plasma is one temperature, the
electron cooling time is $\gsim$ the inflow time of the gas.  As a
result, the accretion flow is always reasonably advection dominated,
because the electrons themselves are.  Therefore, although the
argument for the event horizon in Sgr A* is significantly less
dramatic because of the possibility of substantial mass loss ($\eta_i
\approx 0.01$ for $p \approx 0.8$ vs. $\eta_i \approx 2.6 \times
10^{-5}$ for $p \approx 0$), it nonetheless appears that all viable
ADAF models of Sgr A* have $\eta_i$ substantially smaller than the
usual thin disk value of $\eta_i \approx 0.1$.  This problem should be
investigated in more detail, in particular, with a better treatment of
the dynamics of large $p$ models.

BB98 noted that the presence of winds would cause the difference
between the quiescent luminosities of black hole and neutron star soft
X-ray transients to reduce.  This is discussed further in Menou et
al. (1998).

\subsection{Discussion}

In all of our models we have assumed that winds lead to an $\dot m
\propto r^p$ profile inside $\ro = 10^4$.  This corresponds to
$10^{-4p}$ of the incoming mass being accreted. The spectral models
are, however, primarily sensitive to the fraction of the mass
accreted, rather than to $p$ and $\ro$ separately.  If one favors a
smaller range of radii over which winds are important, this
corresponds roughly to a new $p' = 4p/\log r_w$, where $r_w$ is the
radial extent of the region where winds are important.  Note, however,
that if this approach is taken, our values of $\delta$ may only be
upper limits, since a model with a larger $p' > p$ will rotate more
quickly, and thus have a larger viscous dissipation per unit mass.

In particular, then, our conclusion that low $\delta$ models of Sgr A*
and V404 Cyg are compatible with the observations only for $p \lsim
0.25$ need not conflict with theoretical estimates that $p \sim 1$ is
needed for the Bernoulli parameter of the accreting gas to be
negative.  It may simply mean that winds are important only over $\sim
1-1.5$ decades of radius ($r_w \approx 10-30$), instead of 4 decades
as we have assumed here.  Nonetheless, it is clear that for winds to
be both dynamically crucial (large $p$) and take away the majority of
the mass (large $r_w$), large values of $\delta$ are required.

\noindent{\it Acknowledgments.}  We thank Jeff McClintock, Kristen
Menou, and Jun-Hui Zhao for useful discussions.  Jean-Pierre Lasota
provided a number of useful comments on an earlier version of this
paper.  This work was supported by an NSF Graduate Research Fellowship
and by NASA Grant NAG 5-2837.

\newpage

{
\footnotesize
\StartRef
\noindent {\large \bf References} \\
\Ref Abramowicz, M., Chen, X., Granath, M., \& Lasota, J.P. 1996, ApJ, 471, 762 \\
\Ref Abramowicz, M., Chen, X., Kato, S., Lasota, J.-P., \& Regev, O., 1995,
ApJ, 438, L37 \\
\Ref Backer, D., C, 1982, in Proc. IAU Symposium, eds. D.S. Heeschen \& C. M. Wade,  97, 389 \\
\Ref Blackman, E., 1998, MNRAS in press (astro-ph/ 9710137) \\
\Ref Blandford, R. D. \& Begelman, M. C., 1998, MNRAS submitted (astro-ph/9809083) (BB98) \\
\Ref Blandford, R. D. \& Payne, D.G., 1982, MNRAS, 199, 883 \\
%\Ref Abramowicz, M., Lanza, A. \& Percival, M. J., 1997, ApJ, 479, 179 \\
%\Ref Balbus, S. A., \& Hawley, J.F., 1998, Rev. Mod. Phys., 70, 1 \\
\Ref Bisnovatyi-Kogan, G. S., \& Lovelace R. V. E., 1997, ApJ, 486, L43 \\
\Ref Cannizzo, J. K., 1993, in ``Accretion Disks in Compact Stellar Systems,'' ed. J. Wheeler (Singapore:  World Scientific), p. 6 \\
\Ref Chen, X., Abramowicz, M.A., Lasota, J.-P., Narayan, R., \& Yi, I. 1995, ApJ, 443, L61 \\
\Ref Chen, X., Abramowicz, M.A., Lasota, J.-P., 1997, ApJ, 476, 61 \\

\Ref Coker, R. \& Melia, F., 1997, ApJ, 488, L149 \\

\Ref Di Matteo, T., Fabian, A. C., Rees, M. J., Carilli, C. L., \& Ivison, R. J., 1998, MNRAS, submitted (astro-ph/9807245) (D98)\\
 
%\Ref Duschl, W. J., \& Lesch, H. 1994, A\&A, 286, 431 \\
%\Ref Eckart, A., et al. 1992, Nature, 355, 526 \\
 
\Ref Eckart, A., \& Genzel, R. 1997, MNRAS, 284, 576 \\
 
%\Ref Eckart, A., Genzel, R., Hofmann, R., Sams, B. J., \& Tacconi-Garman, L. E. 1995, ApJ, 445, L23 \\

\Ref Esin, A. A., McClintock, J. E., \& Narayan, R., 1997, ApJ, 489, 86 \\
\Ref Esin, A. A., Narayan, R., Cui, W., Grove, E., \& Zhang, S., 1998, ApJ in press (astro-ph/9711167) \\
\Ref Fabian, A. C. \& Canizares, C. R., 1988, Nature, 333, 829 \\
\Ref Fabian, A. C. \& Rees, M. J., 1995, MNRAS, 277, L55 \\
\Ref Falcke, H., 1996, in IAU Symp. 169, Unsolved Problems of the 
Milky Way, eds. L. Blitz \& P. J. Teuben (Dordrecht: Kluwer), 163 \\

\Ref Falcke, H.,  Goss, W. M., Matsuo, H., Teuben, P., Zhao, J., \& Zylka, R., 1998, ApJ, 499, 731 \\ 
\Ref Gammie, C.F., Narayan, R., \& Blandford, R., 1998, MNRAS submitted (astro-ph/9808036) \\ 
\Ref Gammie, C.F.  \&  Popham, R.G., 1998, ApJ, 498, 313 \\

%\Ref Genzel, R., \& Townes, C. H. 1987, ARAA, 25, 377 \\
\Ref Genzel, R., Eckart, A., Ott, T., \& Eisenhauer, F., 1997, MNRAS, 291, 219 \\
 
\Ref Genzel, R., Hollenbach, D., \& Townes, C. H. 1994, Rep. 
Prog. Phys., 57, 417 \\
 
\Ref Gruzinov, A. 1998, ApJ, 501, 787 \\
\Ref Gruzinov, A. \& Quataert, E., 1998, ApJ submitted (astro-ph/9808278)\\
 
\Ref Haller, J. W., Rieke, M.J., Rieke, G.H., Tamblyn, P., Close, L., \& Melia, F., 1996, 456, 194 (Erratum, 1996, ApJ, 468, 955) \\

%\Ref Haller, J. W., Rieke, M. J., Rieke, G. H., Tamblyn, P., 
%Close, L., \& Melia, F., 1996, 456, 194 (Erratum, 1996, ApJ, 468, 955) \\
 
\Ref Hameury, J. M., Lasota, J. P., McClintock, J. E., \& Narayan, R., 1997, ApJ, 489, 234 \\
 
\Ref Hawley, J. F., Gammie, C. F., \& Balbus, S. A., 1996, ApJ, 464, 690 \\
%\Ref Honma, F., 1996, PASJ, 48, 77 \\
\Ref Ichimaru, S. 1977, ApJ, 214, 840 \\
%\Ref Kato, S., Abramowicz, M., \& Chen, X. 1996, PASJ, 48, 67 \\
\Ref Kato, S., Fukue, J., Mineshige, S., 1998, {\em Black-Hole Accretion Disks}
(Japan: Kyoto University Press) \\
\Ref Lasota, J. P., Abramowicz, M.A., Chen, X., Krolik, J., Narayan, R., \& Yi, I., 1996a, ApJ, 462, 142 \\
\Ref Lasota, J. P., Narayan, R., \& Yi, I., 1996b, A \& A, 314, 813 \\
\Ref Mahadevan, R., 1997, ApJ, 477, 585\\
\Ref Mahadevan, R., 1998, Nature, 394, 651 \\
\Ref Mahadevan, R, Narayan, R., \& Krolik, J. 1997, ApJ,  486, 268 \\
\Ref Mahadevan, R. \& Quataert, E. 1997, ApJ, 490, 605 \\
 
\Ref Manmoto, T., Mineshige, S., \& Kusunose, M., 1997, ApJ, 489, 791\\

\Ref Markoff, S., Melia, F., \& Sarcevic, I., 1997, ApJ, 489, L47 \\ 

\Ref  McClintock, J. E., Horne, K., \& Remillard, R. A., 1995, ApJ, 442, 35 \\

\Ref Melia, F. 1992, ApJ, 387, L25 \\
\Ref Menou, K., Esin, A. A., Narayan, R., Garcia, Lasota, J.P., \& M., McClintock, J. E., 1998, in preparation \\
\Ref Menou, K., Quataert, E., \& Narayan, R., 1998, in Proc. of the 8th Marcel
Grossmann Meeting on General Relativity (astro-ph/9712015) \\
\Ref Menten, K. M., Reid, M. J., Eckart, A., \& Genzel, R. 1997, ApJ, 475, L111 \\
\Ref Nakamura, K. E., Masaaki, K., Matsumoto, R., \& Kato, S. 1997, PASJ, 49, 503 \\
\Ref Narayan, R. 1996, ApJ, 462, 136 \\
\Ref Narayan, R., Barret, D., \& McClintock, J. E., 1997a, ApJ, 482, 448\\
\Ref Narayan, R., Kato, S., \& Honma, F. 1997, ApJ, 476, 49 \\
\Ref Narayan, R., Mahadevan, R., Grindlay, J.E., Popham, R.G., \& Gammie, C., 1998a, ApJ, 492, 554 \\
\Ref Narayan, R., Mahadevan, R., \& Quataert, E., 1998b, in {\em The Theory
of Black Hole Accretion Discs}, eds. M.A. Abramowicz, G. Bjornsson, and J.E. Pringle (Cambridge:  Cambridge University Press) (astro-ph/9803131) \\
\Ref Narayan, R., McClintock, J.E., \& Yi, I., 1996, ApJ, 457 821 \\
\Ref Narayan, R., \& Yi, I., 1994, ApJ, 428, L13 \\
\Ref Narayan, R., \& Yi, I., 1995, ApJ, 444, 231 \\ 
\Ref Narayan, R., Yi, I., \& Mahadevan, R., 1995, Nature, 374, 623 \\
\Ref Peitz, J. \& Appl, S. 1997, MNRAS, 286, 681 \\
\Ref Quataert, E. 1998, ApJ, 500, 978 \\
\Ref Quataert, E. \& Gruzinov, A., 1998, ApJ submitted (astro-ph/9803112) \\
\Ref Quataert, E. \& Narayan, R., 1998, ApJ submitted (astro-ph/9810117)\\
 
\Ref Rees, M. J., 1982, in {\em The Galactic Center} (ed. G. R. Riegler
\& R. D. Blandford). AIP, p. 166, New York \\
 
\Ref Rees, M. J., Begelman, M. C., Blandford, R. D., \& Phinney,
E. S., 1982, Nature, 295, 17 \\
  
\Ref Rogers, A. E. E., et al., 1994, ApJ, 434, L59 \\
%\Ref Rybicki, G. \& Lightman, A., 1969, { \em Radiative Processes in Astrophysics} (New York:  John Wiley \& Sons) \\

\Ref Serabyn, E., Carlstrom, J., Lay, O., Lid, D.C., Hunter, T.R., \& Lacy, J.H., 1997, ApJ, 490, L77 \\

\Ref Shahbaz, T., Ringwald, F.A., Bunn, J.C., Narylor, T., Charles, P.A., \& Casares, J., 1994, MNRAS, 271, L10 \\
 
%\Ref Shakura, N. I., \& Sunyaev, R. A., 1973, A\&A, 24, 337 \\
 
%\Ref Shapiro, S. L., Lightman, A. P., \& Eardley, D. M. 1976, ApJ, 204, 187 \\

\Ref Vargas. M. et al., 1996, in {\em The Galactic Center}, ed. R. Gredel, 431 \\
\Ref Wheeler, J. C., 1996,  in {\em Relativistic Astrophysics}, eds. B. Jones \& D. Markovic (Cambridge: Cambridge University press) \\

}

\newpage

\begin{deluxetable}{lcccccccc}
%\doublespace
%\footnotesize
\tablecaption{Model Parameters for V404 Cyg:  $\ro = 10^4$, 
$m = 12$, $\dot M_{\rm in} \equiv \dot M_{\rm out} 10^{-4 p}$} 
\tablewidth{0pt}
\tablehead{
\colhead{Fig.}         & 
\colhead{$\alpha$}        &
\colhead{$\beta$}        &
\colhead{$\delta$}          &
\colhead{$p$}          &
\colhead{$\mo \times 10^2$}	&
\colhead{$T_{\rm e, max} \times 10^{-10}$}	&
\colhead{$L/{\dot M_{\rm out}} c^2$} 	&
\colhead{$L/{\dot M_{\rm in}} c^2$}}
\startdata
1a & 0.1 & 10 & 0.01 & 0 & 0.1 & 1.0 & $1.1 \times 10^{-3}$ & $1.1 \times 10^{-3}$ \nl 
1a & 0.1 & 10 & 0.01 & 0.2 & 0.54 & 0.92 & $1.6 \times 10^{-4}$ & $1.0 \times 10^{-3}$ \nl  
1a & 0.1 & 10 & 0.01 & 0.4 & 1.6 & 0.87 & $2.2 \times 10^{-5}$ & $8.8 \times 10^{-4}$ \nl  
1a & 0.1 & 10 & 0.01 & 0.6 & 2.0 & 0.6 & $7.5 \times 10^{-6}$ & $2.0 \times 10^{-3}$ \nl  
\nl
1b & 0.1 & 1 & 0.01 & 0.4 & 1.2 & 0.57 & $6.0 \times 10^{-5}$ & $2.4 \times 10^{-3}$ \nl  
1b & 0.1 & 10 & 0.01 & 0.4 & 1.6 & 0.87 & $2.2 \times 10^{-5}$ & $8.8 \tfo $ \nl  
1b & 0.1 & 100 & 0.01 & 0.4 & 1.7 & 1.0 & $1.2 \times 10^{-5}$ & $4.8 \tfo $\nl  
\nl
2a & 0.3 & 10 & 0.01 & 0.4 & 2.6 & 0.89 & $1.9 \times 10^{-5}$ & $7.6 \tfo$ \nl  
2a & 0.1 & 10 & 0.01 & 0.4 & 1.6 & 0.87 & $2.2 \times 10^{-5}$ & $8.8 \tfo$ \nl  
2a & 0.03 & 10 & 0.01 & 0.4 & 0.59 & 0.77 & $5.5 \times 10^{-5}$ & $2.2 \tt$ \nl  
\nl
2b & 0.1 & 10 & $10^{-3}$ & 0.4 & 1.6 & 0.87 & $2.2 \times 10^{-5}$ & $8.8 \tfo$ \nl  
2b & 0.1 & 10 & 0.01 & 0.4 & 1.6 & 0.87 & $2.2 \times 10^{-5}$ & $ 8.8 \tfo$ \nl  
2b & 0.1 & 10 & 0.03 & 0.4 & 1.2 & 0.94 & $3.6 \times 10^{-5}$ & $ 1.4 \tt$ \nl  
2b & 0.1 & 10 & 0.1 & 0.4 & 1.0 & 1.15 & $6.2 \times 10^{-5}$ & $ 2.5 \tt$ \nl  
2b & 0.1 & 10 & 0.3 & 0.4 & 0.58 & 1.8 & $1.7 \times 10^{-4}$ & $6.8 \tt$ \nl  
\nl
3a & 0.3 & 1 & 0.01 & 0 & 0.1 & 0.74 & $1.8 \times 10^{-3}$ & $1.8 \tt$ \nl 
3a & 0.1 & 10 & 0.01 & 0 & 0.1 & 1.0 & $1.1 \times 10^{-3}$ & $1.1 \tt$\nl 
3a & 0.03 & 30 & 0.01 & 0 & 0.068 & 1.1 & $1.1 \times 10^{-3}$ & $1.1 \tt$ \nl 
\nl
3b & 0.1 & 10 & 0.01 & 0 & 0.1 & 1.0 & $1.1 \times 10^{-3}$ & $1.1 \tt$ \nl 
3b & 0.1 & 10 & 0.1 & 0 & 0.05 & 1.3 & $2.5 \times 10^{-3}$ & $2.5 \tt$ \nl 
3b & 0.1 & 10 & 0.3 & 0 & 0.01 & 2.4 & $6.4 \times 10^{-3}$ & $6.4 \tt$ \nl 
\nl
8a & 0.1 & 10 & 0.01 & 0 & 0.1 & 1.0 & $1.1 \times 10^{-3}$ & $1.1 \tt$ \nl 
8a & 0.1 & 10 & 0.3 & 0.4 & 0.58 & 1.8 & $1.7 \times 10^{-4}$ & $6.8 \tt$ \nl  
8a & 0.1 & 10 & 0.75 & 0.8 & 0.64 & 4.8 & $6.3 \times 10^{-5}$ & $0.1$ \nl  
\enddata
\label{tab-v}
\end{deluxetable}

\newpage

\begin{deluxetable}{lcccccccc}
%\doublespace
%\footnotesize
\tablecaption{Model Parameters for Sgr A*:  $\ro = 10^4$, 
$m = 2.5 \times 10^6$, $\dot M_{\rm in} \equiv \dot M_{\rm out} 10^{-4 p}$} 
\tablewidth{0pt}
\tablehead{
\colhead{Fig.}         & 
\colhead{$\alpha$}        &
\colhead{$\beta$}        &
\colhead{$\delta$}          &
\colhead{$p$}          &
\colhead{$\mo \times 10^4$}	&
\colhead{$T_{\rm e, max} \times 10^{-10}$}	&
\colhead{$L/{\dot M_{\rm out}} c^2$} 	&
\colhead{$L/{\dot M_{\rm in}} c^2$}}
\startdata
4a & 0.1 & 10 & 0.01 & 0 & 0.68 & 2.0 & $2.6 \times 10^{-5}$ & $2.6 \tfi$\nl 
4a & 0.1 & 10 & 0.01 & 0.2 & 1.8 & 1.1 & $5.5 \times 10^{-7}$ & $3.5 \tsi$\nl  
4a & 0.1 & 10 & 0.01 & 0.4 & 2.4 & 0.63 & $1.2 \times 10^{-7}$ &$4.8 \tsi$\nl  
4a & 0.1 & 10 & 0.01 & 0.6 & 2.8 & 0.37 & $8.1 \times 10^{-8}$ & $2.0 \tfi$ \nl  
\nl
4b & 0.1 & 1 & 0.01 & 0.4 & 1.9 & 0.47 & $1.2 \times 10^{-7}$ & $4.8 \tsi$ \nl 
4b & 0.1 & 10 & 0.01 & 0.4 & 2.4 & 0.63 & $1.2 \times 10^{-7}$ & $4.8 \tsi$ \nl
4b & 0.1 & 100 & 0.01 & 0.4 & 2.4 & 0.65 & $1.2 \times 10^{-7}$ & $4.8 \tsi$ \nl
\nl
5a & 0.3 & 10 & 0.01 & 0.4 & 4.0 & 0.72 & $ 7.6 \times 10^{-8}$ & $3.1\tsi$\nl 
5a & 0.1 & 10 & 0.01 & 0.4 & 2.4 & 0.63 & $1.2 \times 10^{-7}$ &$4.8\tsi$\nl  
5a & 0.03 & 10 & 0.01 & 0.4 & 1.0 & 0.54 & $2.8 \times 10^{-7}$ &$1.1 \tfi$\nl  
\nl
5b & 0.1 & 10 & 0.01 & 0.4 & 2.4 & 0.63 & $1.2 \times 10^{-7}$ &$4.8\tsi$\nl  
5b & 0.1 & 10 & 0.1 & 0.4 & 1.7 & 1.6 & $ 3.1 \times 10^{-7}$ &$1.2\tfi$\nl 
5b & 0.1 & 10 & 0.3 & 0.4 & 1.7 & 3.5 & $3.1 \times 10^{-6}$ &$1.2\tfo$\nl  
\nl
6a & 0.3 & 1 & 0.01 & 0 & 0.69 & 1.6 & $7.6 \times 10^{-5}$ &$7.6\tfi$\nl  
6a & 0.1 & 10 & 0.01 & 0 & 0.68 & 2.0 & $2.6 \times 10^{-5}$ &$2.6\tfi$\nl  
6a & 0.03 & 30 & 0.01 & 0 & 0.48 & 1.6 & $7.7 \times 10^{-6}$ &$7.7\tsi$\nl  
\nl
6b & 0.1 & 10 & 0.01 & 0 & 0.68 & 2.0 & $2.6 \times 10^{-5}$ &$2.6\tfi$\nl  
6b & 0.1 & 10 & 0.1 & 0 & 0.54 & 2.3 & $2.6 \times 10^{-5}$ &$2.6\tfi$\nl  
6b & 0.1 & 10 & 0.3 & 0 & 0.11 & 5 & $1.0 \times 10^{-4}$ &$1.0\tfo$\nl  
\nl
8b & 0.1 & 10 & 0.01 & 0 & 0.68 & 2.0 & $2.6 \times 10^{-5}$ &$2.6\tfi$\nl 
8b & 0.1 & 10 & 0.4 & 0.4 & 1.6 & 4.4 & $6.9 \times 10^{-6}$ &$2.7\tfo$\nl  
8b & 0.1 & 10 & 0.75 & 0.8 & 1.2 & 13.0 & $3.7 \times 10^{-6}$ &$5.9\tt$\nl  
\enddata
\label{tab-s}
\end{deluxetable}

\newpage

\begin{deluxetable}{lcccccccc}
%\doublespace
%\footnotesize
\tablecaption{Model Parameters for NGC 4649:  $\ro = 10^4$, 
$m = 8 \times 10^9$, $\dot M_{\rm in} \equiv \dot M_{\rm out} 10^{-4 p}$} 
\tablewidth{0pt}
\tablehead{
\colhead{Fig.}         & 
\colhead{$\alpha$}        &
\colhead{$\beta$}        &
\colhead{$\delta$}          &
\colhead{$p$}          &
\colhead{$\mo$}	&
\colhead{$T_{\rm e, max} \times 10^{-10}$}	&
\colhead{$L/{\dot M_{\rm out}} c^2$} 	&
\colhead{$L/{\dot M_{\rm in}} c^2$}}
\startdata
7a & 0.1 & 10 & 0.01 & 0 & $10^{-3}$ & 1.9 & $3.1 \times 10^{-4}$&$3.1\tfo$ \nl 
7a & 0.1 & 10 & 0.3 & 0.25 & $10^{-3}$ & 3.4 & $3.6 \times 10^{-5}$&$3.6\tfo$ \nl  
7a & 0.1 & 10 & 0.01 & 0.25 &  $10^{-3}$ & 1.0 & $2.1 \times 10^{-6}$&$2.1\tfi$ \nl  
7a & 0.1 & 10 & 0.3 & 0.54 & $10^{-3}$ & 2.4 & $6.6 \times 10^{-7}$&$9.5\tfi$ \nl  
\nl
7b & 0.1 & 10 & 0.01 & 0 & $10^{-4.5}$ & 1.9 & $2.6 \times 10^{-6}$&$2.6\tsi$ \nl 
7b & 0.1 & 10 & 0.3 & 0.25 &$10^{-4.5}$ & 3.5 & $9.1 \times 10^{-7}$&$9.1\tsi$ \nl  
7b & 0.1 & 10 & 0.01 & 0.25 & $10^{-4.5}$ & 1.0 & $5.7 \times 10^{-8}$&$5.7\tse$ \nl  

\enddata
\label{tab-ngc}
\end{deluxetable}

\newpage
\vskip 5in
\newpage

\begin{figure}
\plottwo{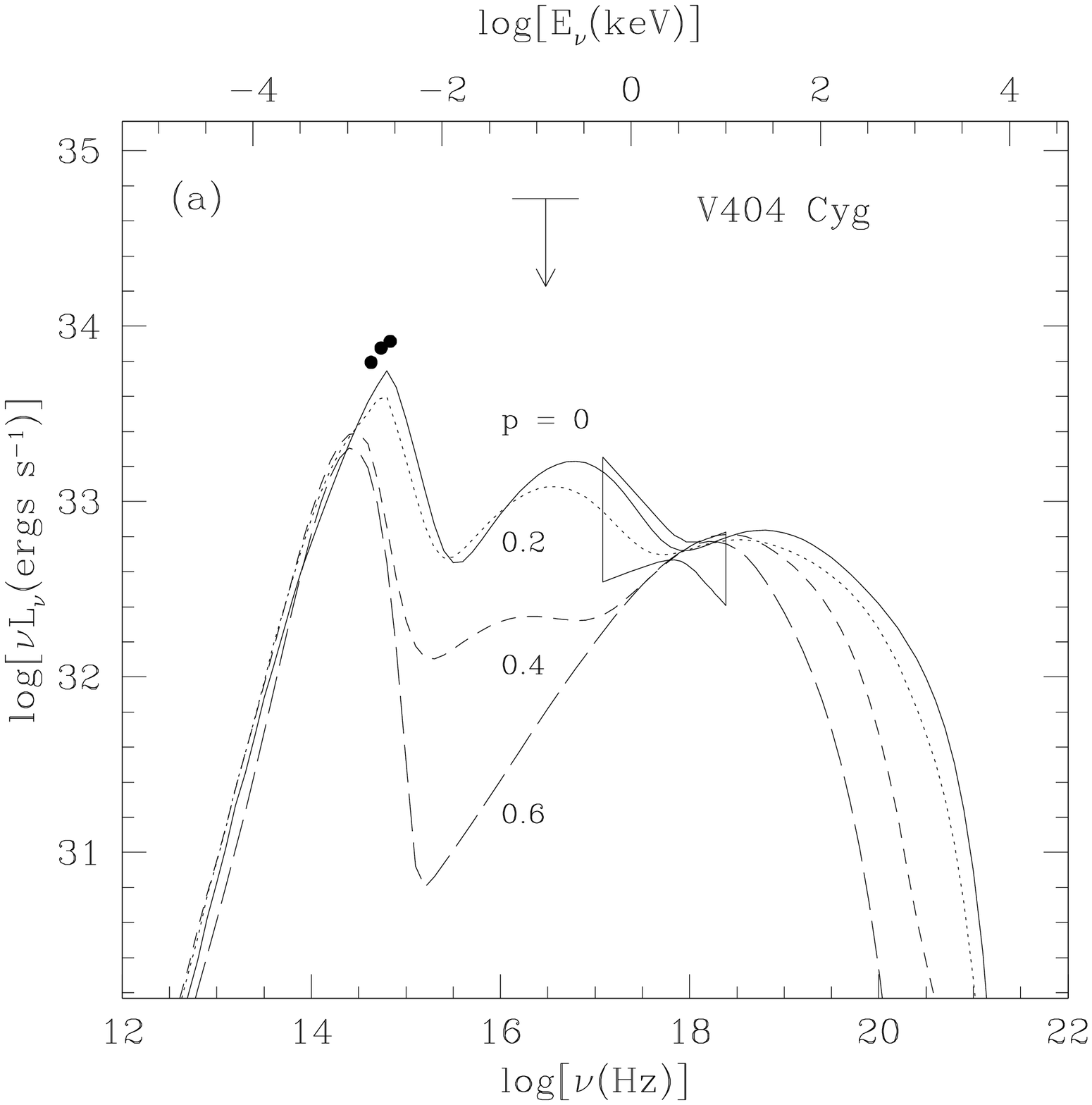}{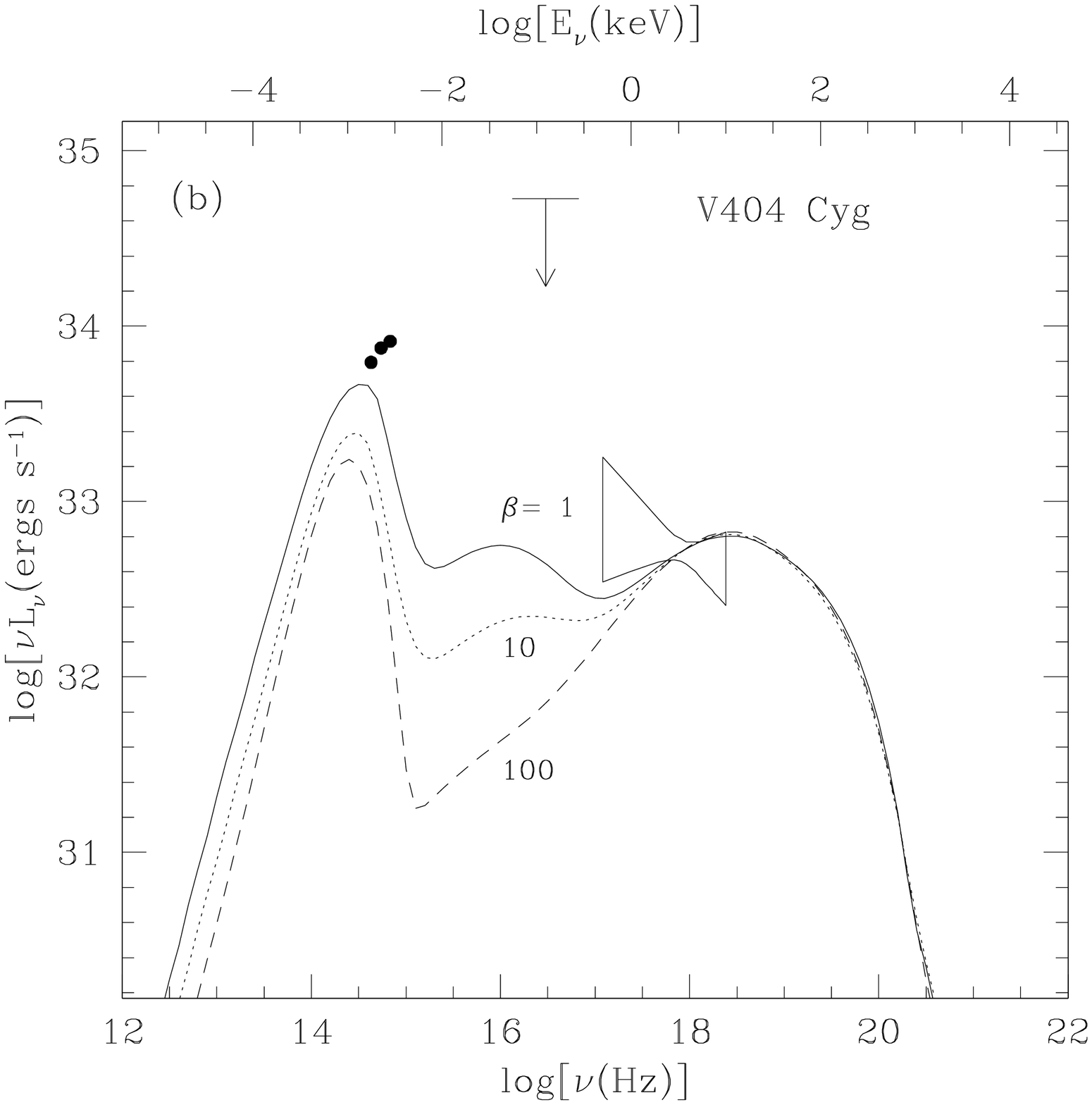}
\caption{(a) Spectral models of V404 Cyg for several values of $p$,
taking $\alpha = 0.1$, $\beta = 10$, and $\delta = 0.01$.  (b) Models
for several $\beta$, taking $\alpha = 0.1$, $p = 0.4$, and $\delta =
0.01$.}
\end{figure}

\begin{figure}
\plottwo{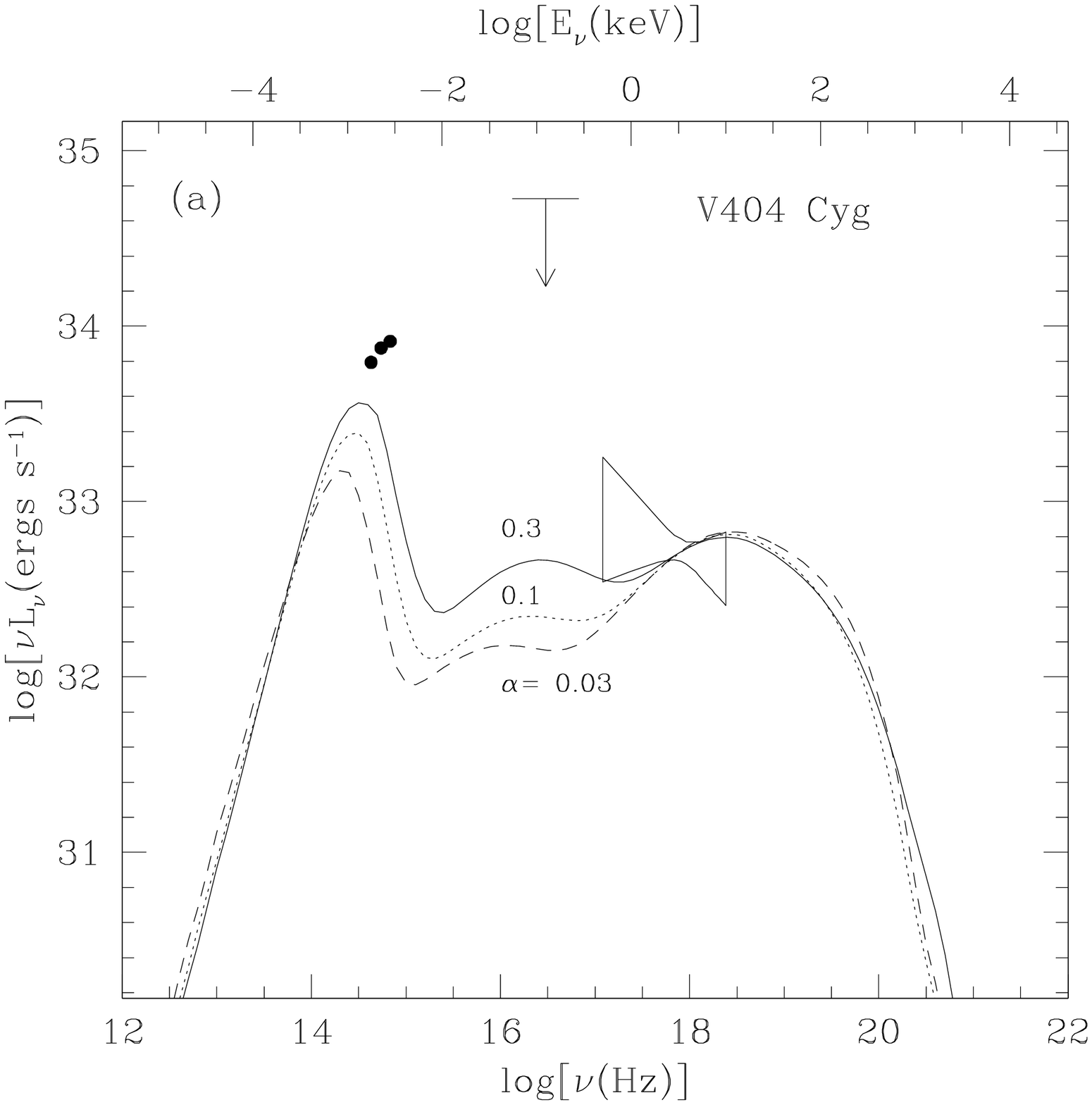}{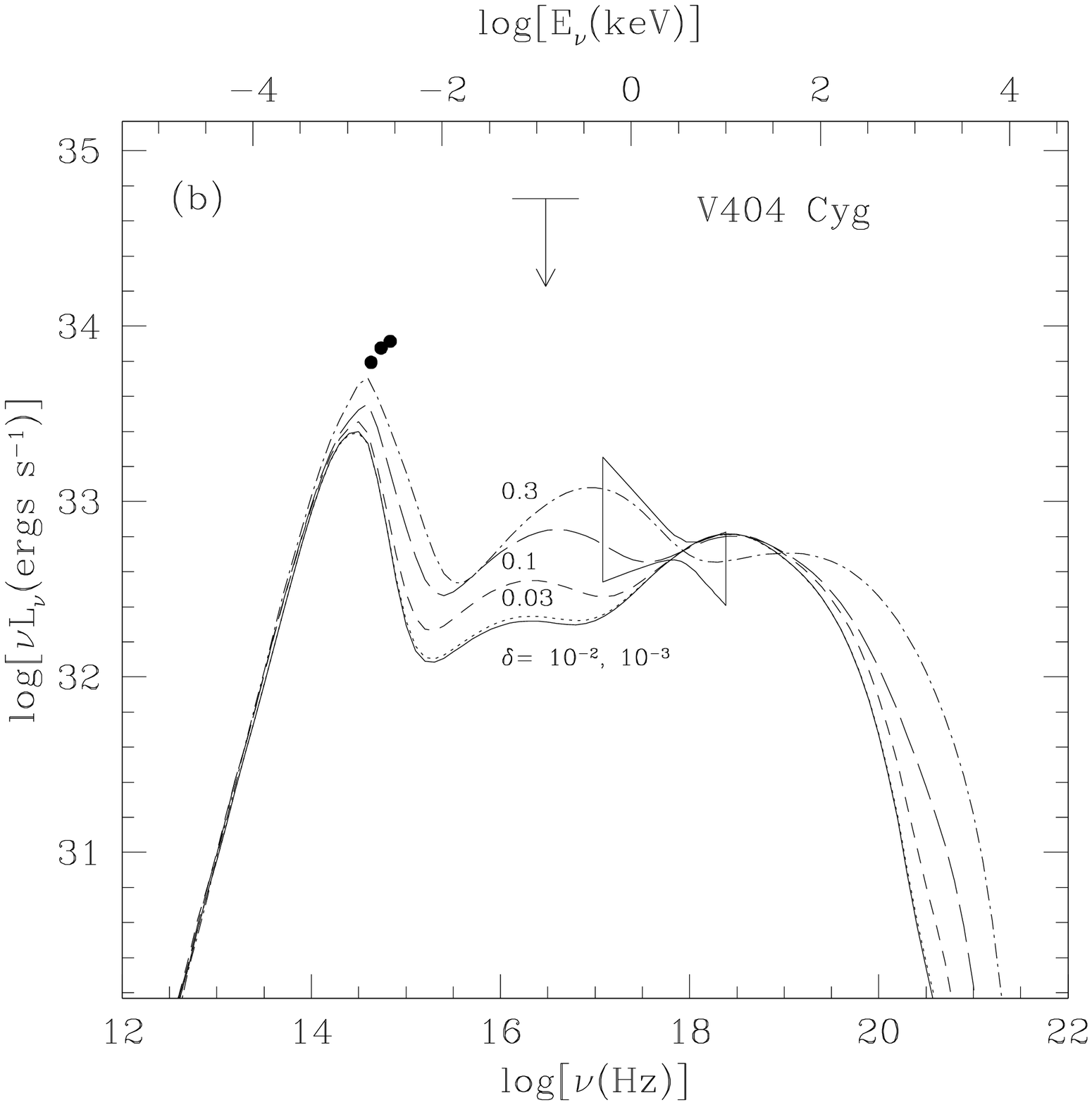}
\caption{(a) Spectral models of V404 Cyg for several values of $\alpha$,
taking $p = 0.4$, $\beta = 10$, and $\delta = 0.01$.  (b) Models for
several $\delta$, taking $\alpha = 0.1$, $p = 0.4$, $\beta = 10$.}
\end{figure}

\newpage

\begin{figure}
\plottwo{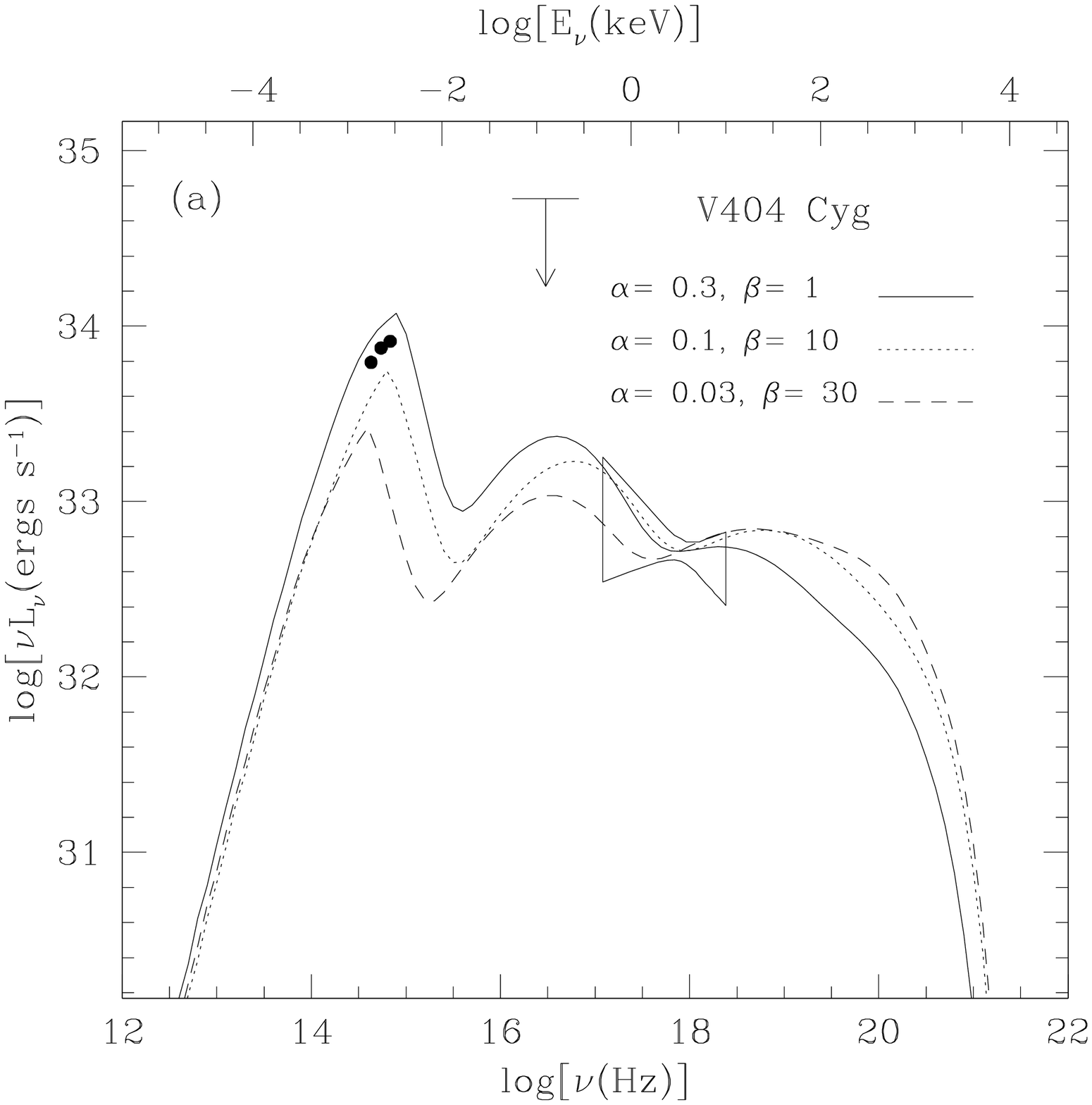}{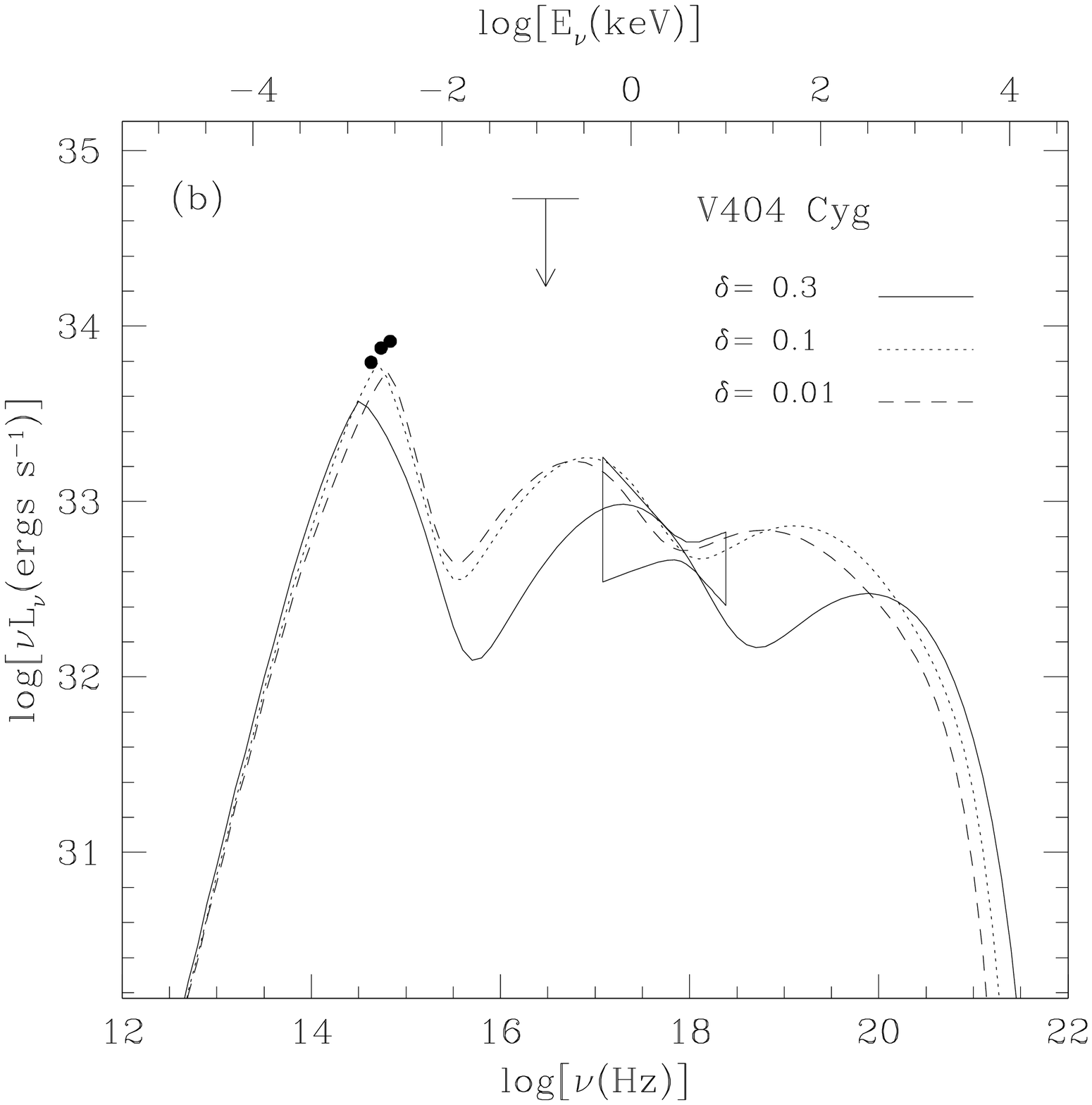}
\caption{(a) No wind ($p = 0$) spectral models of V404 Cyg for several
values of $\alpha$ and $\beta$, taking $\delta = 0.01$. (b) Models for
several values of $\delta$, taking $p = 0$, $\alpha = 0.1$, and $\beta =
10$.}
\end{figure}

\begin{figure}
\plottwo{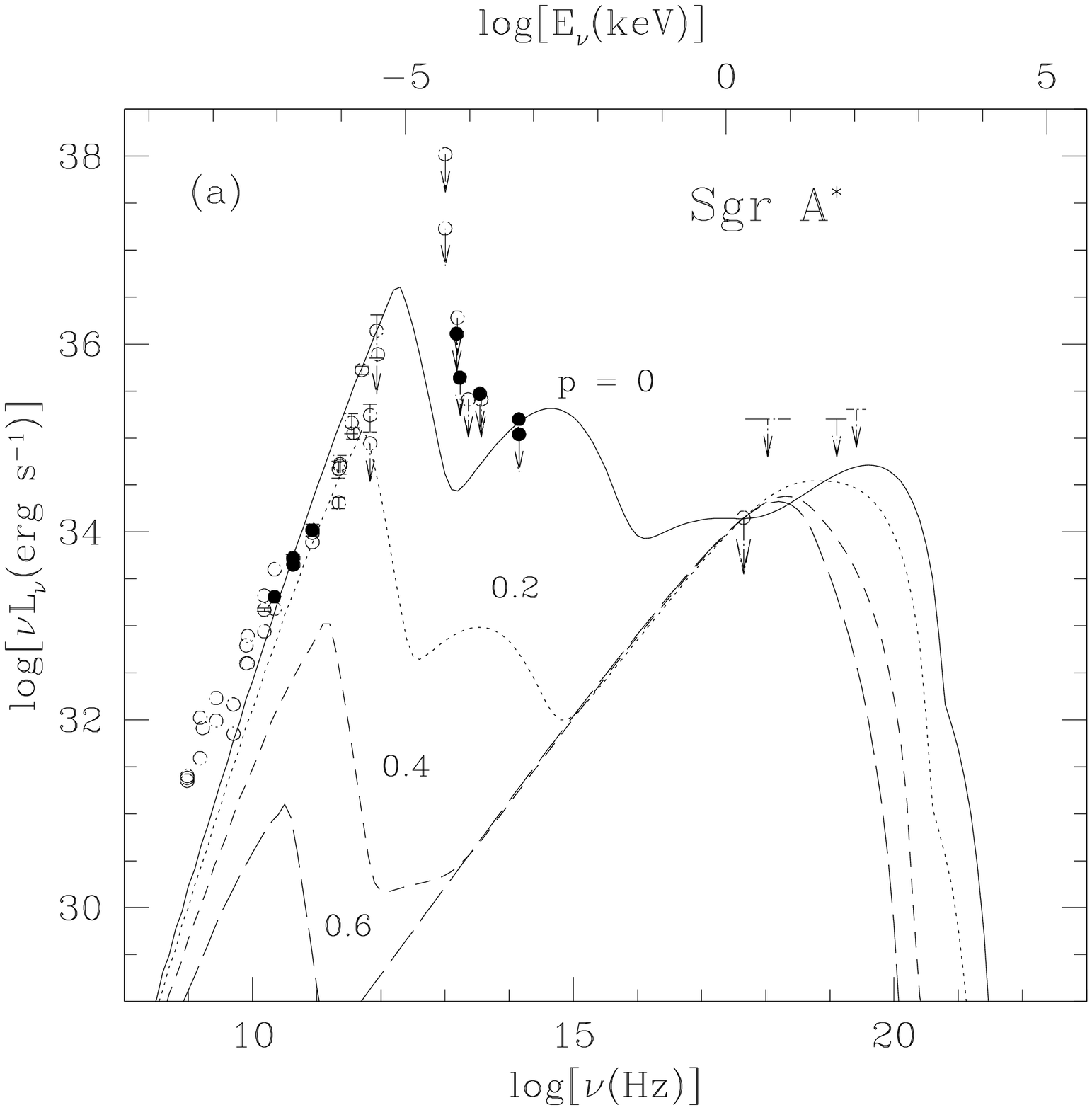}{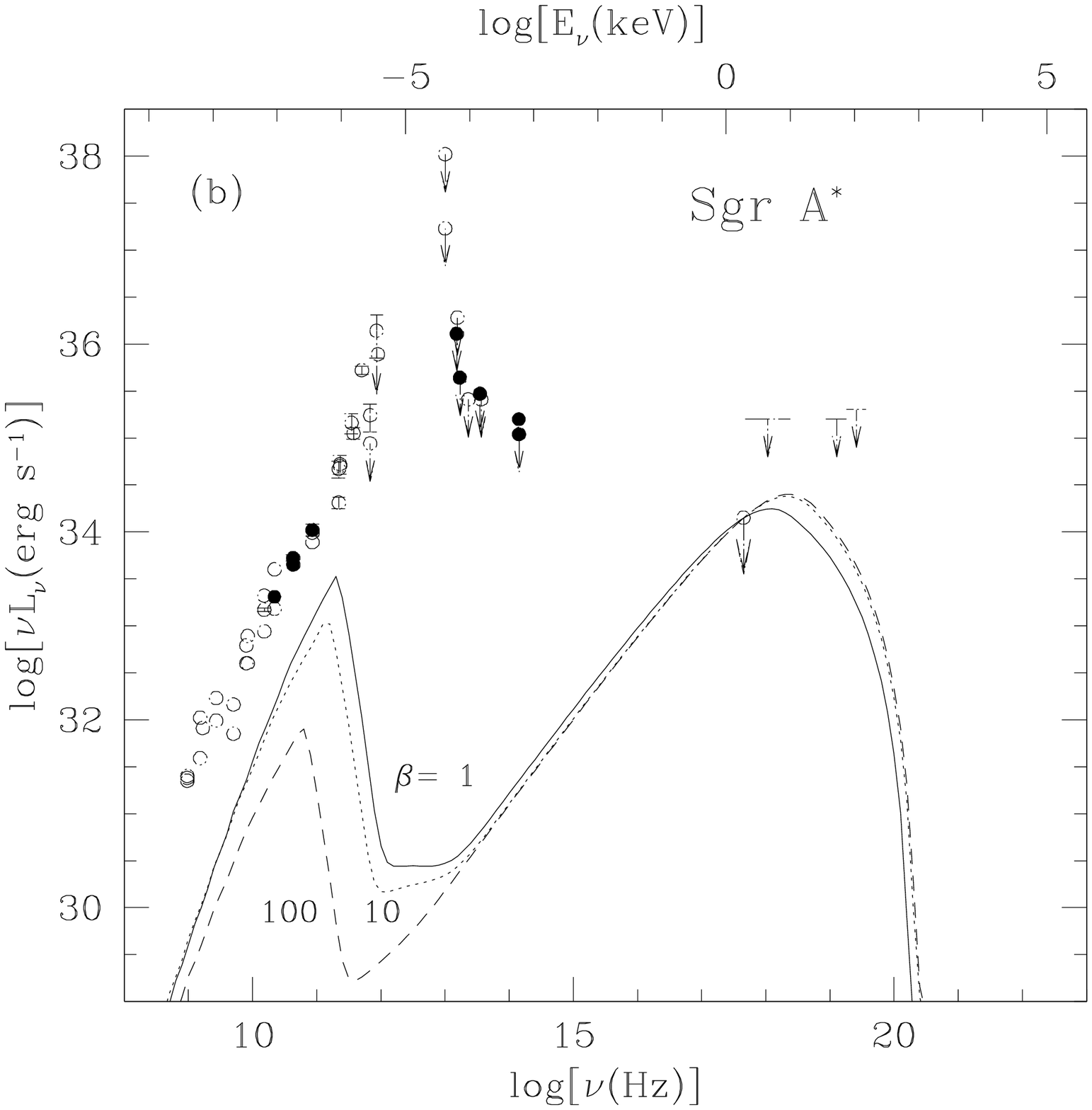}
\caption{(a) Spectral models of Sgr A* for several values of $p$,
taking $\alpha = 0.1$, $\beta = 10$, and $\delta = 0.01$.  (b) Models
for several $\beta$, taking $\alpha = 0.1$, $p = 0.4$, and $\delta =
0.01$.}
\end{figure}

\newpage

\begin{figure}
\plottwo{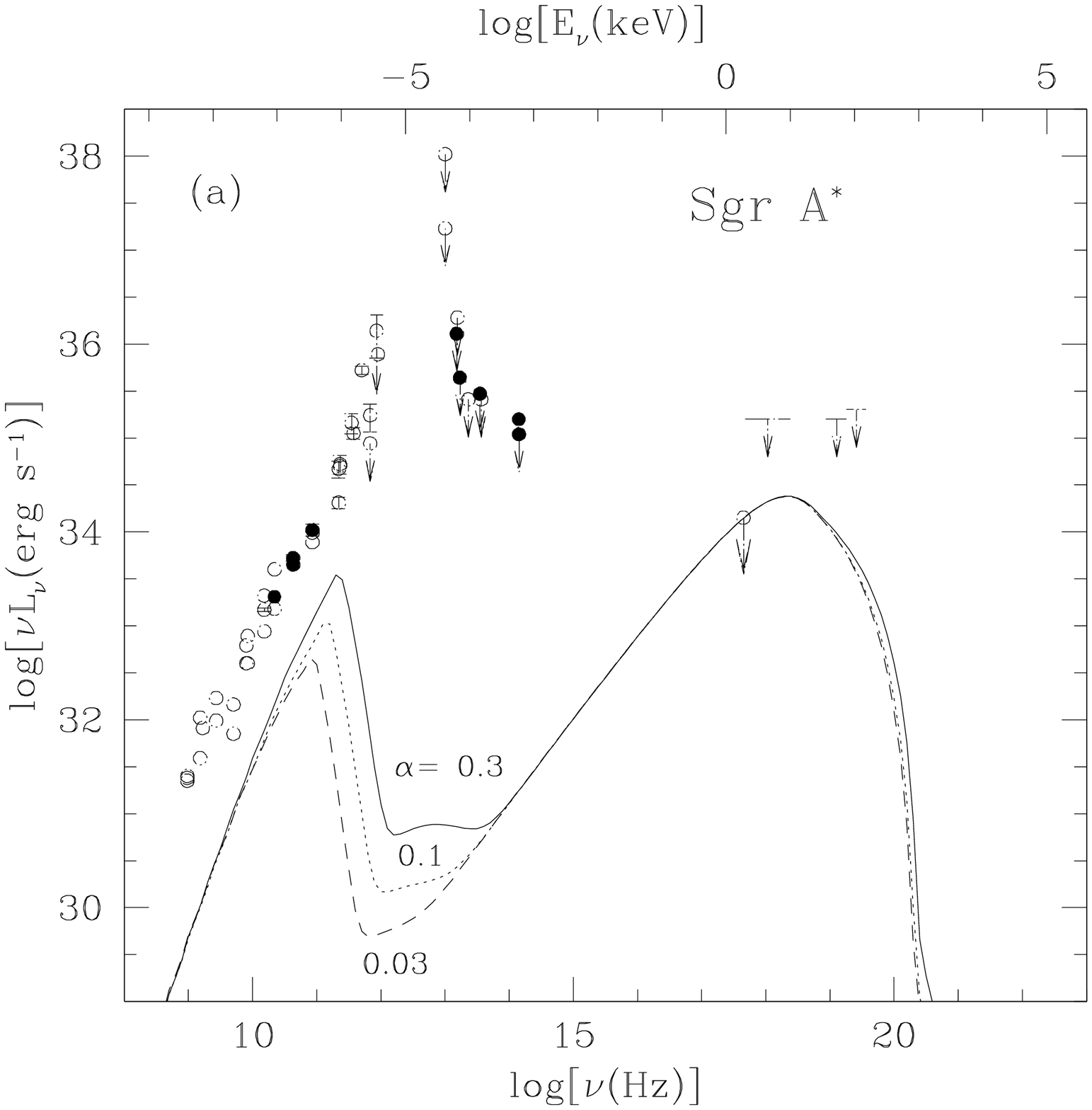}{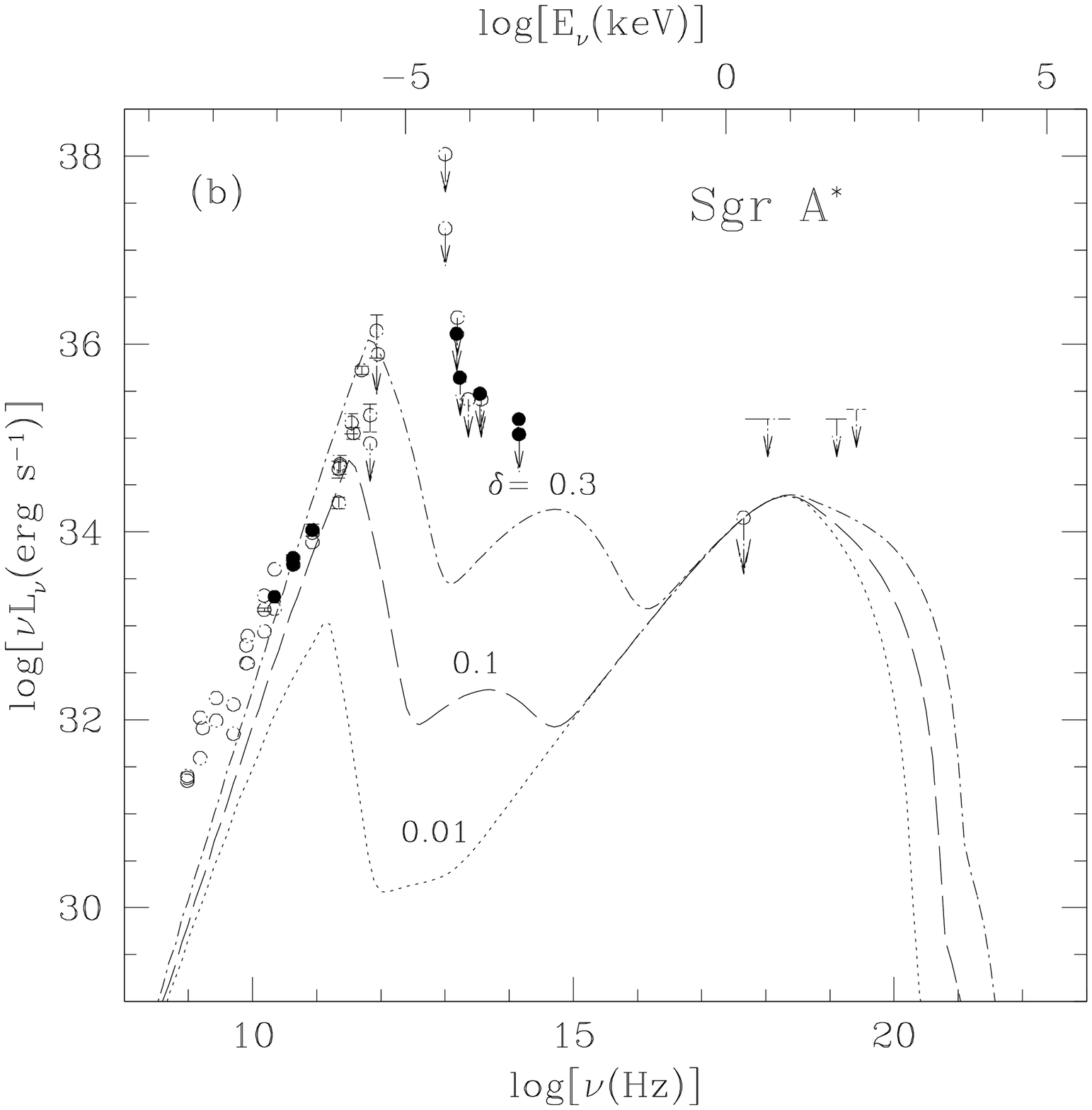}
\caption{(a) Spectral models of Sgr A* for several values of $\alpha$,
taking $p = 0.4$, $\beta = 10$, and $\delta = 0.01$.  (b) Models for
several $\delta$, taking $\alpha = 0.1$, $p = 0.4$, $\beta = 10$.}
\end{figure}

\begin{figure}
\plottwo{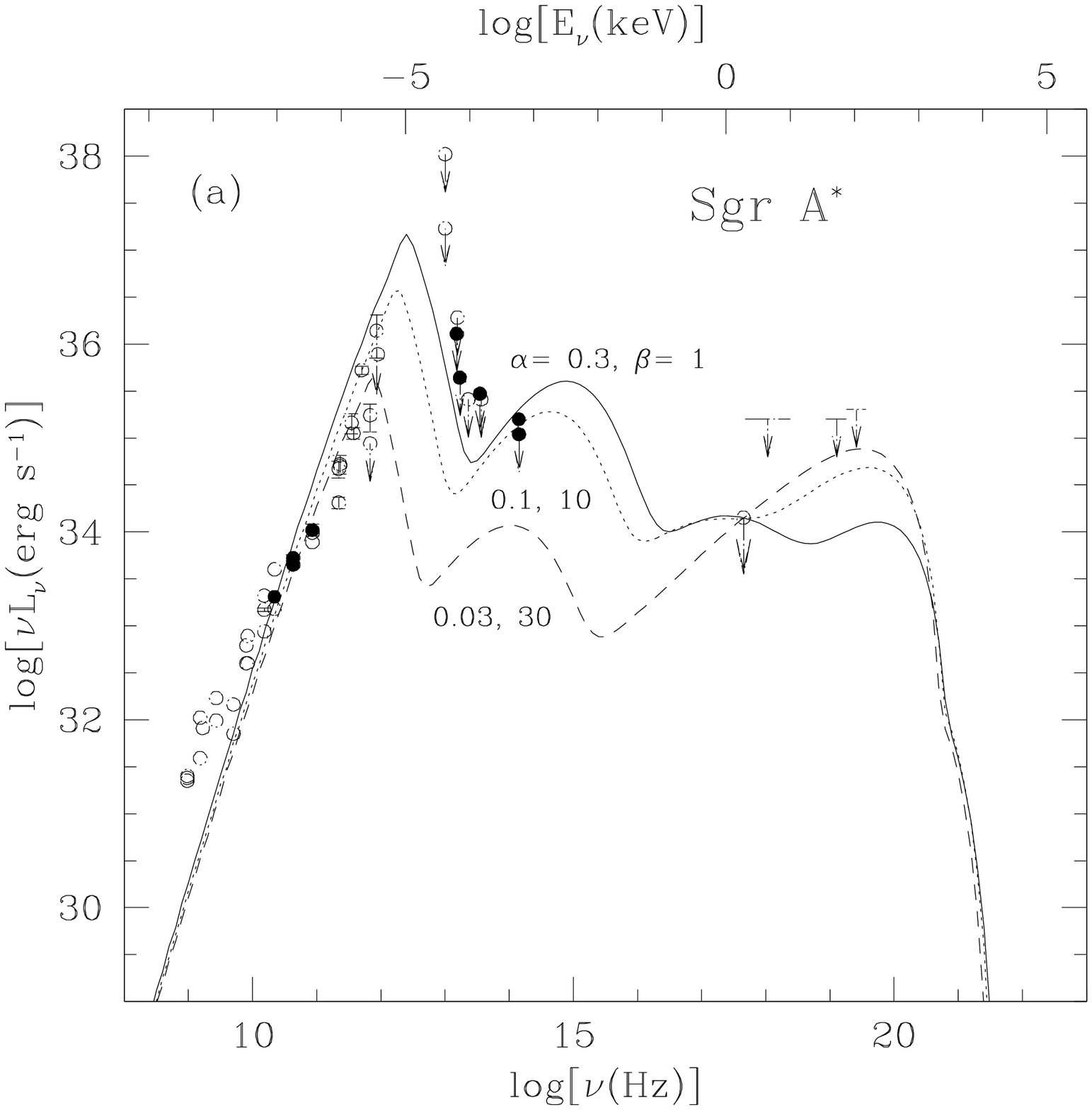}{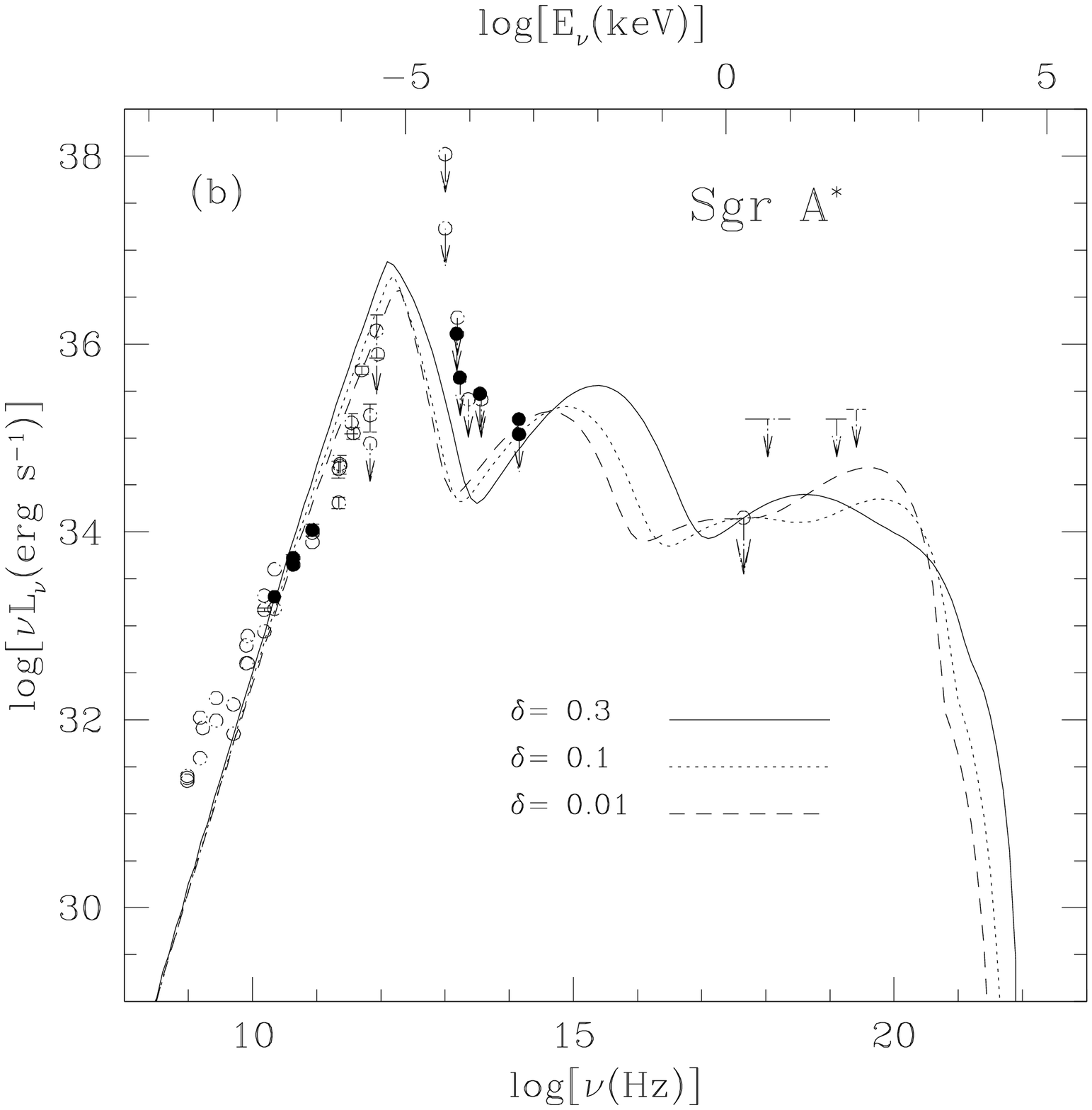}
\caption{(a) Spectral models of Sgr A* for several values of $\alpha$
and $\beta$, taking $p = 0$, $\delta = 0.01$.  (b) Models for several
$\delta$, taking $\alpha = 0.1$, $\beta = 10$, and $p = 0$.}
\end{figure}

\newpage

\begin{figure}
\plottwo{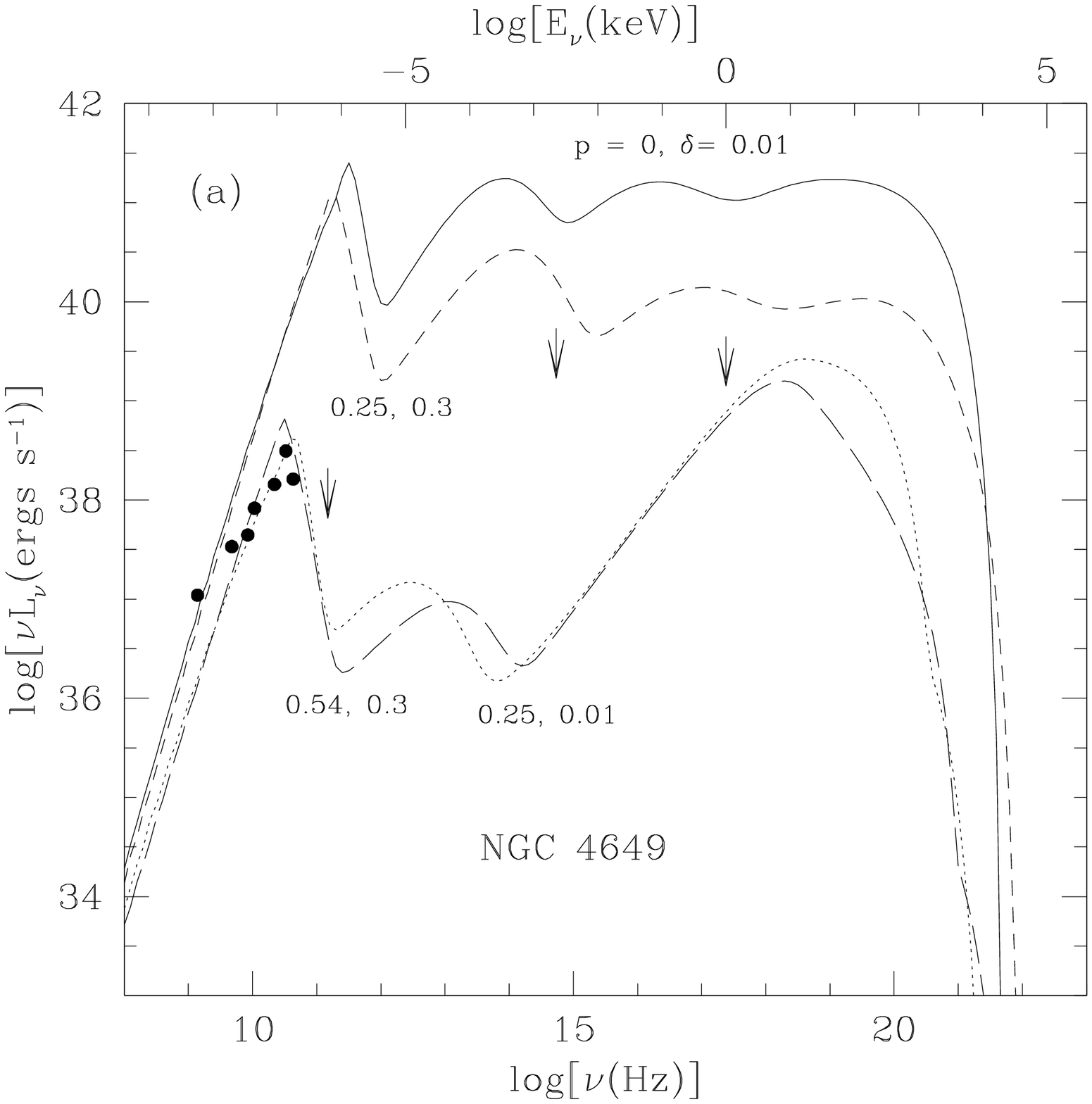}{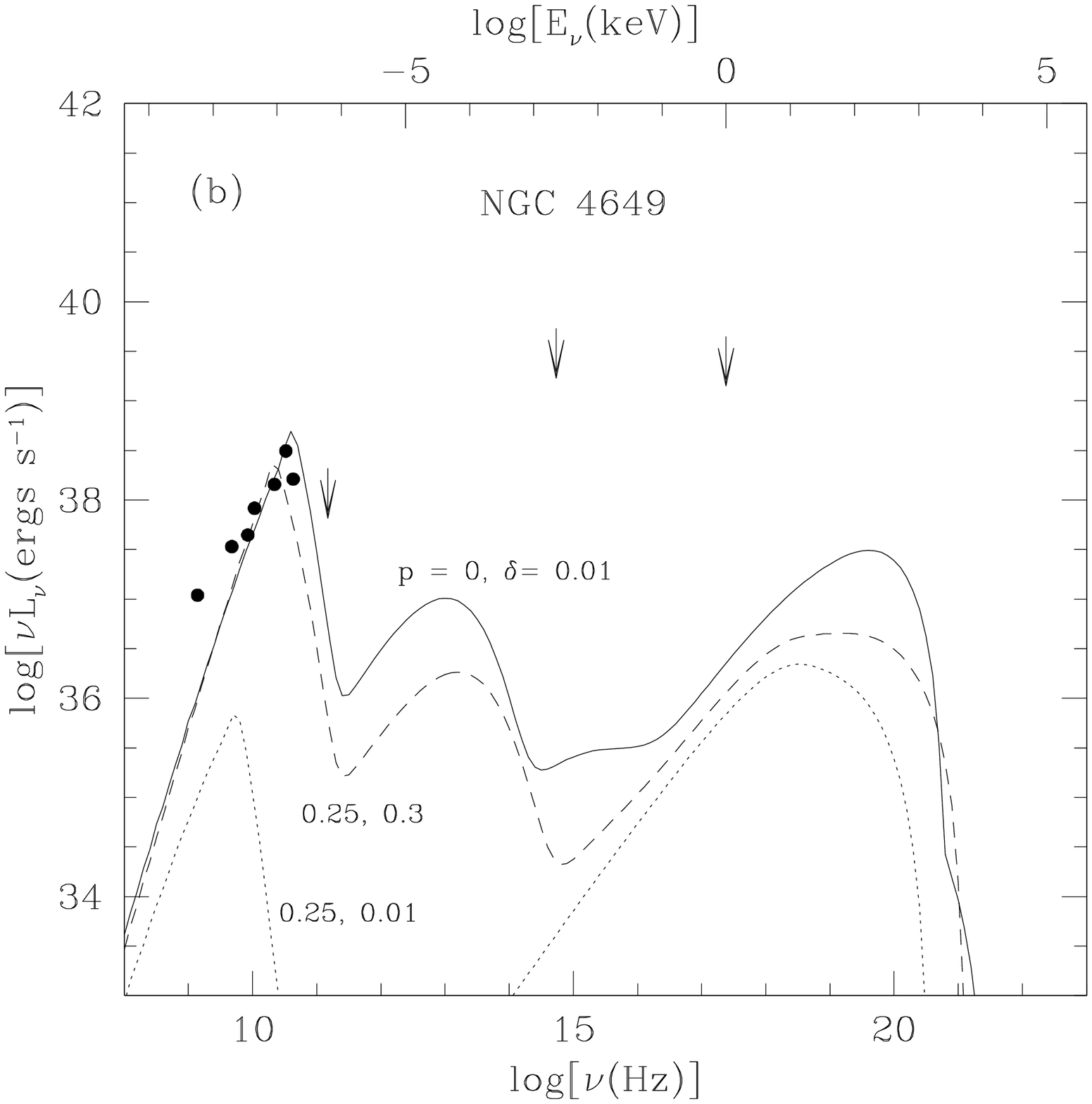}
\caption{Spectral models of NGC 4649 for several values of $p$ and
$\delta$, taking $\alpha = 0.1$, $\beta = 10$, $\ro = 10^4$, and $m =
8 \times 10^9$.  Panel (a) assumes that the accretion rate at the
outer edge of the flow is $\mo = 10^{-3}$, while panel (b) takes $\mo
= 10^{-4.5}$.}
\end{figure}

\begin{figure}
\plottwo{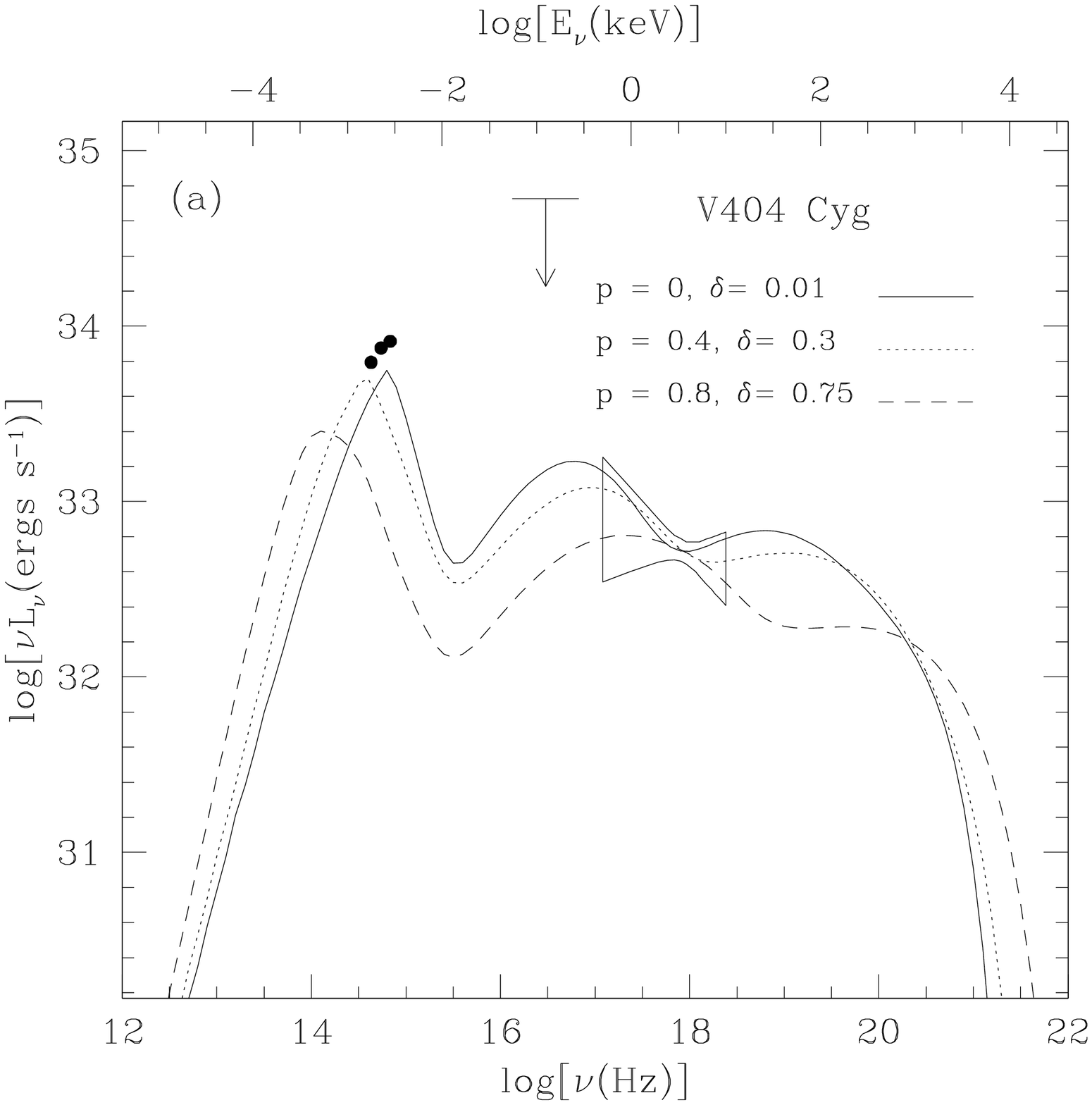}{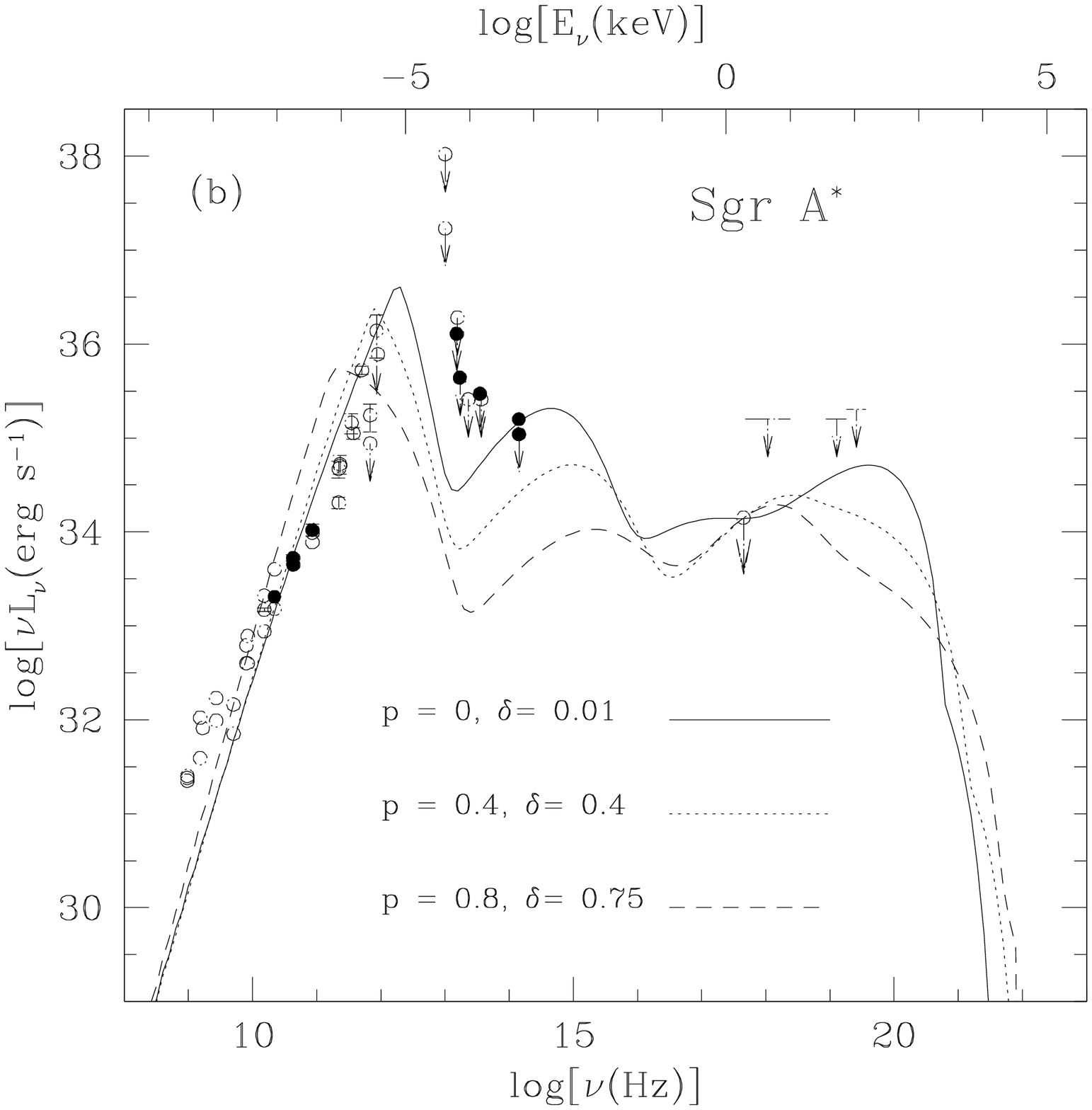}
\caption{Spectral models of V404 Cyg (Fig. 8a) and Sgr A* (Fig. 8b)
for several values of $p$ and $\delta$, taking $\alpha = 0.1$, $\beta
= 10$, and $\ro = 10^4$.}
\end{figure}

\end{document}